\documentclass[aps,prd,preprint,superscriptaddress]{revtex4-1}
\usepackage{amsmath,amssymb}
\usepackage{graphicx}
\makeatletter

\@addtoreset{equation}{section}
\makeatother
\def\vector#1{\mbox{\boldmath ${#1}$}} %
\def\m#1{\mathrm{#1}} 
\def\be#1{\begin{equation}#1\end{equation}} 

\def\pdif#1#2{\frac{\partial {#1}}{\partial {#2}}} 

\def\exp#1{\mathrm{e}^{#1}} 

\def\bra#1{\langle{#1}|} 
\def\cket#1{|{#1}\rangle} 

\def\summation#1#2{\sum^{#2}_{#1}} 

\begin{document}
\preprint{KEK-CP-345}

\title{
  Short-distance charmonium correlator on the lattice with
  M\"obius domain-wall fermion and a determination of charm quark mass
}


\author{Katsumasa Nakayama}
\email[]{katumasa@post.kek.jp}
\affiliation{Department of Physics, Nagoya University, Nagoya, 464-8602, Japan}
\affiliation{KEK Theory Center, High Energy Accelerator Research Organization (KEK), Tsukuba 305-0801, Japan}

\author{Brendan Fahy}
\affiliation{KEK Theory Center, High Energy Accelerator Research Organization (KEK), Tsukuba 305-0801, Japan}

\author{Shoji Hashimoto}
\affiliation{KEK Theory Center, High Energy Accelerator Research Organization (KEK), Tsukuba 305-0801, Japan}
\affiliation{School of High Energy Accelerator Science, The Graduate University for Advanced Studies (Sokendai),Tsukuba 305-0801, Japan
}

\collaboration{JLQCD Collaboration}
%
%
\date{\today}
\begin{abstract}
  We calculate charmonium correlators on the lattice with 
  2+1-flavors of sea quarks and charm valence quark both described by
  the M\"obius domain-wall fermion.
  Temporal moments of the correlators are calculated and matched to
  perturbative QCD formulae to extract the charm quark mass $m_c(\mu)$
  and strong coupling constant $\alpha_s(\mu)$.
  Lattice data at three lattice spacings, 0.044, 0.055, and
  0.080~fm, are extrapolated to the continuum limit.
  The correlators in the vector channel are confirmed to be consistent
  with the experimental data for $e^+e^-\to c\bar{c}$,
  while the pseudo-scalar channel is used to extract $m_c(\mu)$ and
  $\alpha_s(\mu)$.
  We obtain 
  $m_\m{c}(3 \m{\ GeV})$ = 1.003(10)~GeV and 
  $\alpha_s^{\overline{\m{MS}}(4)}(3\m{\ GeV})$ = 0.253(13).
  Dominant source of the error is the truncation of perturbative
  expansion at $\alpha_s^3$.
\end{abstract}

\maketitle
\section{Introduction}
\label{sec:introduction}
Numerical simulation of lattice QCD offers non-perturbative
calculation of correlation functions on the Euclidean lattice.
While one usually uses the long-distance correlators to extract the 
mass and matrix elements of hadrons, 
the same correlators at short distances also provide a rich source of information.
The vector current correlator, for instance, may be used to test QCD
by comparing the lattice calculation with the experimental data
available for the $R$ ratio 
$\sigma_{e^+e^-\to q\bar{q}}/\sigma_{e^+e^-\to\mu^+\mu^-}$.
The correlator becomes mostly perturbative at high energy scales, but
the non-perturbative effect is still important.
Another important use of the short-distance regime is the application
of perturbation theory, from which one can extract the fundamental
parameters such as the strong coupling constant $\alpha_s$ and charm
quark mass $m_c$.

The HPQCD and Karlsruhe collaboration used the pseudo-scalar
charmonium correlator to achieve a precise determination of 
$m_c$ and $\alpha_s$ \cite{Allison:2008xk}, which has further been
improved and extended to include the determination of the bottom quark
mass \cite{McNeile:2010ji,Chakraborty:2014aca}.
The basic idea is to use a perturbative QCD calculation performed at
the order of $\alpha_s^3$ to express temporal moments of the charmonium
correlator calculated non-perturbatively on the lattice.
Since the perturbative expansion is given as a function of $\alpha_s$
and $m_c$, one can solve the equations to determine these parameters.
The precision achieved is among the best for these important
fundamental parameters of QCD.

In this work we utilize the same method to extract $m_c$ and
$\alpha_s$. 
Our lattice data are independent from those used by the
HPQCD collaboration.
We use the lattice ensembles generated with 2+1 flavors of light sea
quarks described by the M\"obius domain-wall fermion formulation \cite{Brower:2012vk}.
The valence charm quark is also treated by the same fermion
formulation.
Discretization effects expected for relatively large charm
quark mass compared to the lattice spacing are largely removed by
extrapolating to the continuum limit using the data at three lattice
spacings, $a\simeq$ 0.080, 0.055, and 0.044~fm.
The light quark masses in the simulations are in the range
corresponding to the pion mass of 230--500~MeV, which do not cover the
physical value but their effect on the charmonium correlator is
minor.

On the perturbative side, we use the same perturbative coefficients as
those in the previous works
\cite{Allison:2008xk,McNeile:2010ji,Chakraborty:2014aca}.
We estimate the truncation error by examining the dependence on the
renormalization scale $\mu_\alpha$ to define the coupling constant
$\alpha_s(\mu_\alpha)$ as well as that on $\mu_m$ that defines the running
charm quark mass $m_c(\mu_m)$ appearing in the perturbative expansion.

Our results are in reasonable agreement with those of 
\cite{Allison:2008xk,McNeile:2010ji,Chakraborty:2014aca}.
The estimated error is slightly larger, 
because of different systematic effect as well as different error
estimates. 
We also try to validate the lattice calculation by providing a
comparison to the experimental data available for the vector channel
through the $R$-ratio.
It mainly serves as a test of the discretization effects, 
which is an important source of the systematic error for heavy
quarks. 
We find that the continuum extrapolation is nearly flat,
confirming that the discretization error for charm quark is well under
control in our setup.

This paper is organized as follows.
In Section~\ref{sec:time_moments} we review the method of 
\cite{Allison:2008xk,McNeile:2010ji,Chakraborty:2014aca}
as well as the formulae to compare
the temporal moments with the experimental data.
Some details of our lattice calculation are given in
Section~\ref{sec:lattice_details}.
Lattice results for the vector current correlator 
and the comparison with the experimental data are given in
Section~\ref{sec:vector_moments}, which is followed by corresponding
results for pseudo-scalar correlator in
Section~\ref{sec:pseudoscalar_moments}.
The issues in the matching to perturbative results and its possible
uncertainties are discussed in Section~\ref{sec:matching},
and results for charm quark mass and strong coupling constant are
finally given in Section~\ref{sec:determination}.
Our conclusions are in Section~\ref{sec:conclusion}.

\section{Charmonium correlators and their temporal moments}
\label{sec:time_moments}

\subsection{Charmonium correlators}
We calculate the pseudo-scalar and vector charmonium correlators with
vanishing spatial momentum
\begin{eqnarray}
  \label{eq:G^PS}
  G^{PS}(t) & = & a^6 \sum_{\vector{x}} (am_c)^2 
  \langle 0| j_5 (\vector{x},t)j_5 (0,0) |0\rangle,
  \\
  \label{eq:G^V}
  G^{V}(t) & = & \frac{a^6}{3}\sum_{k=1}^3 \sum_{\vector{x}} 
  Z_V^2
  \langle 0| j_k (\vector{x},t)j_k (0,0) |0\rangle,
\end{eqnarray}
on the lattice.
The currents are defined as
$j_5 = i\bar{\psi_c}\gamma_5\psi_c$
and 
$j_k = \bar{\psi_c}\gamma_k\psi_c$
with charm quark field $\psi_c$ on the lattice.
Given the factor $a^6$, both $G^{PS}(t)$ are $G^V(t)$ are
dimensionless.
The pseudo-scalar density operator $j_5$ is multiplied by a (bare)
charm quark mass $m_c$ such that the correlator becomes
renormalization scale invariant, while a possible renormalization
factor $Z_V$ for the vector current $j_k$ defined on the lattice is explicitly multiplied in
(\ref{eq:G^V}).

We then construct the temporal moments as
\begin{eqnarray}
  \label{eq:momentPS}
  G_n^{PS} & = & \sum_t \left(\frac{t}{a}\right)^n G^{PS}(t),
  \\
  \label{eq:momentV}
  G_n^V & = & \sum_t \left(\frac{t}{a}\right)^n G^V(t),
\end{eqnarray}
with $n$ an even integer equal to or larger than four.
(The correlator $\langle 0|j(x)j(0)|0\rangle$ diverges as $1/|x|^6$ in
the small-separation limit,
and the lower moments contain ultraviolet divergences.)
On the lattice, the time coordinate $t/a$ runs between $-T/2a+1$ and
$T/2a$ with $T$ the temporal extent of the lattice.

\begin{figure}[tbp]
  \centering
  \includegraphics[width=10.0cm,angle=-90]{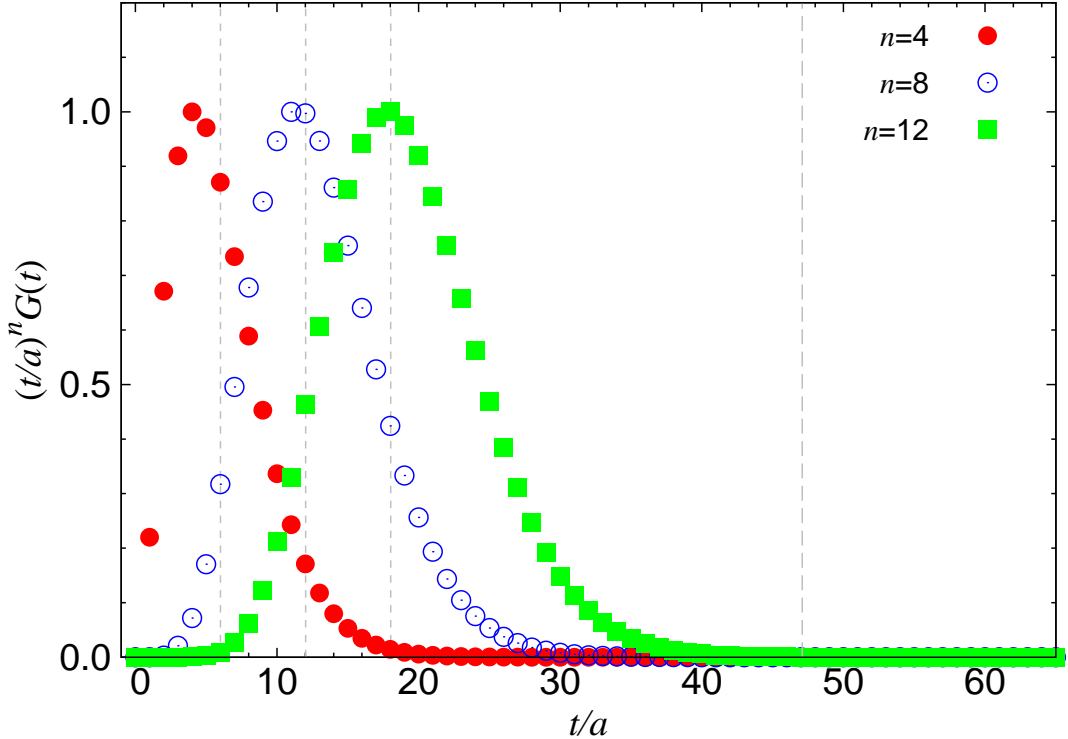}
  \caption{
    $(t/a)^n G(t)$ on the lattice of size $64^3\times 128$ at $a$ =
    0.044~fm.
    The function is normalized by its peak.
    The data for $n$ = 4 (filled circle), 8 (open circle), and 12
    (square) are shown.
    The long-dashed line around $t/a\sim48$ represents the point of 
    $\pi$/($300$ MeV), which is the distance that non-perturbative effect
    dominates. 
    Three vertical dashed lines show the position of peak $n/M$ 
    ($n$ = 4, 8 and 12) for the single exponential function
    $\exp{-Mt}$.
    Here, $Ma$ = 0.6656.  
  }
  \label{fig:window}
\end{figure}

Since the charmonium correlators $G^{PS}(t)$ and $G^V(t)$ decay
exponentially at large $t$ by the mass of the corresponding lowest
energy states $\eta_c$ and $J/\psi$, respectively, the temporal moments
(\ref{eq:momentPS}) and (\ref{eq:momentV}) are sensitive only to the
relatively short-range correlations.
For an exponential function $e^{-Mt}$, where $M$ represents the mass
of $\eta_c$ or $J/\psi$, the largest contribution to the
$n$-th moment comes from the region of $t\sim n/M$.
In the presence of excited state contributions, the dominant region is 
slightly shifted to smaller $t$'s.
Figure~\ref{fig:window} illustrates typical examples of the integrand
$(t/a)^nG^{PS}(t)$ to construct the $n$-th moments.
Lattice data at $a$ = 0.044~fm are taken and data for $n$ = 4, 8,
and 12 are shown.
The lowest moment, $n=4$, receives a significant contribution from small
$t$ range, $(t/a)\simeq$ 1--2, where the discretization effect could be
substantial.
For higher moments $n$ = 8 and 12, the sum is not affected much by
the small $t$ range.

The vector correlator and its moments may be related to those in the
continuum theory and to the experimental data.
The vacuum polarization function $\Pi^V(q^2)$ is defined through
\begin{equation}
  (q^\mu q^\nu-q^2g^{\mu\nu})\Pi^V(q^2)
  = i\int d^4x\, e^{iqx}
  \bra{0}j^\mu(x) j^\nu(0)\cket{0}.
\end{equation}
Derivatives of $\Pi^V(q^2)$ with respect to $q^2$,
\begin{equation}
  \label{eq:g_n}
  g_{2k+2}^V
  = 
  (2m(\mu))^{2k}
  \frac{12\pi^2Q_f^2}{k!}
  \left(\frac{\partial}{\partial q^2}\right)^k
  \left(\Pi^V(q^2)\right)|_{q^2=0},
\end{equation}
may be related to the experimental data for the
$e^+e^-\to c\bar{c}$ process, 
{\it i.e.} the $R$-ratio
$R(s)\equiv\sigma_{e^+e^-\to c\bar{c}}(s)/\sigma_{e^+e^-\to\mu^+\mu^-}(s)$,
as
\begin{equation}
  \frac{12\pi^2Q_f^2}{k!}\left(\frac{\partial}{\partial q^2}\right)^k
  \left(\Pi(q^2)\right)|_{q^2=Q_0^2} 
  = 
  M_k
  \equiv
  \int_{s_0}^{\infty} ds \frac{1}{(s-Q_0 ^2)^{k+1}} R(s).
\end{equation}
Here $Q_f$ stands for the electromagnetic charge of charm quark.
The lower end of the integral $s_0$ should be set below the $J/\psi$
mass.
The reference scale $Q_0^2$ is arbitrary but is often taken at
$Q_0^2=0$.
Using this notation we may write the relation between the temporal moments
on the lattice and the observable as
\begin{equation}
  \label{eq:G_n^V}
  G_n^V = \frac{g_n^V}{(am(\mu))^{n-2}}.
\end{equation}
A direct comparison of the lattice results with the 
experimental values for $M_n$ (or their
phenomenological estimates)
is given in Section~\ref{sec:vector_moments}.
The phenomenological estimates of $M_n$ can be found in \cite{Chetyrkin:2006xg,Boughezal:2006px,Kuhn:2007vp,Maier:2009fz}.

For the pseudo-scalar density correlator
\begin{equation}
  q^2 \Pi^{PS} (q^2) = i \int d^4x\, e^{iqx} 
  \langle 0| j_5(x)j_5(0)|0\rangle,
\end{equation}
there is no such experimental information available, while
the relation between the temporal moments and the derivatives of
the vacuum polarization function may be written as
\begin{equation}
  \label{eq:G=g}
  G_n^{PS} = \frac{g_n^{PS}}{(am(\mu))^{n-4}}
\end{equation}
with $g_n^{PS}$ analogously defined as in (\ref{eq:g_n}).

The continuum vacuum polarization functions 
can be parametrized as
\begin{eqnarray}
  \Pi^{PS} (q^2) 
  & = & 
  \frac{3}{16\pi^2} \sum_{k=-1}^\infty C_k^{PS} z^k,
  \\
  \Pi^V(q^2) 
  & = &
  \frac{3}{16\pi^2} \sum_{k=-1}^\infty C_k^V z^k,
\end{eqnarray}
with $z=q^2/(2m_c(\mu))^2$.
In perturbation theory, the coefficients $C_k^{PS}$ and $C_k^V$ are
expanded in terms of $\alpha_s(\mu)/\pi$:
\begin{eqnarray}
  C_k &=& C_k^{(0)} + \frac{\alpha_s(\mu)}{\pi} 
  \left( C_k^{(10)}+ C_k^{(11)}l_{m}\right) 
  \nonumber\\
  & & + \left( \frac{\alpha_s(\mu)}{\pi}\right) ^2 
  \left( C_k^{(20)} + C_k^{(21)}l_{m} + C_k^{(22)}l_{m}^2 \right) 
  \nonumber\\
  & & + \left( \frac{\alpha_s(\mu)}{\pi}\right)^3 
  \left( C_k^{(30)} + C_k^{(31)}l_{m} + C_k^{(32)}l_{m}^2 +
    C_k^{(33)}l_{m} ^3 \right) + ...
\end{eqnarray}
with $l_m = \log{(m^2_c(\mu)/\mu^2)}$. 
(Here, $C_k$ and its expansion coefficients are those of either
$C_k^{PS}$ or $C_k^V$.)
The perturbative calculation has been performed up to
$O(\alpha_s^3)$
\cite{McNeile:2010ji,Chetyrkin:2006xg,Boughezal:2006px,%
Maier:2007yn,Hoang:2008qy,Maier:2009fz,Kiyo:2009gb,Greynat:2010kx,Greynat:2011zp}.
The calculation is conventionally performed in the $\overline{\m{MS}}$
renormalization scheme, and the coupling constant $\alpha_s(\mu)$ and
running quark mass $m_c(\mu)$ are given in that scheme at
a renormalization scale $\mu$.
The relevant coefficients for $n_f$ = 4 are summarized in Table~\ref{tab:ck} for
convenience. 
%

\begin{table}
\begin{tabular}{cc|cccccccccc} 
  \hline
  $n$ & $k$  
  &$C^{(0)}_k$ &$C^{(10)}_k$ &$C^{(11)}_k$ &$C^{(20)}_k$
  &$C^{(21)}_k$ &$C^{(22)}_k$ &$C^{(30)}_k$ &$C^{(31)}_k$
  &$C^{(32)}_k$ &$C^{(33)}_k$ 
  \\ \hline\hline
  4&1  
  &1.33333  &3.11111    &0.00000    &0.115353    &-6.48148    &0.00000    &-1.22241   &2.50084    &13.5031    &0.00000 \\
  6&2
  &0.533333  &2.06420    &1.06667    &7.23618    &1.590947    &-0.0444444            &7.06593    &-7.58522    &0.550549  &0.0320988\\ 
  8&3  
&0.304762  &1.21171    &1.21905    &5.99920    &4.33726    &1.16825    &14.5789    &7.36258    &4.25232    &-0.0649030\\
  10&4  
&0.203275  &0.712756    &1.21905    &4.26701    &4.80644    &2.38730    &13.3285    &14.7645    &11.0345    &1.45891\\
  12&5  
&0.1478    &0.4013    &1.1821    &2.9149    &4.3282    &3.4971    &    &16.0798    &16.6772    &4.4685\\
  14&6  
&0.1137    &0.1944    &1.1366    &1.9656    &3.4173    &4.4992    &    &14.1098    &19.9049    &8.7485\\
  16&7  
&0.0909    &0.0500    &1.0912    &1.3353    &2.2995    &5.4104    &    &10.7755    &20.3500    &14.1272\\
  18&8  
&0.0749    &-0.0545    &1.0484    &0.9453    &1.0837    &6.2466   &    &7.2863    &17.9597    &20.4750\\
 \hline
\end{tabular}
\caption{
  Perturbative coefficients for the pseudo-scalar correlator.
  The results for $n_f$ = 4 are summarized from \cite{Maier:2007yn,Maier:2009fz}.
}
\label{tab:ck}
\end{table}

\begin{table}
\begin{tabular}{cc|cccccccccc} 
  \hline
  $n$ & $k$  
  &$C^{(0)}_k$ &$C^{(10)}_k$ &$C^{(11)}_k$ &$C^{(20)}_k$
  &$C^{(21)}_k$ &$C^{(22)}_k$ &$C^{(30)}_k$ &$C^{(31)}_k$
  &$C^{(32)}_k$ &$C^{(33)}_k$ 
  \\ \hline\hline
    4&1  
  &1.06667   & 2.55473   & 2.13333   & 2.49671   & 3.31303   & -0.0888889   & -5.64043   & 4.06686   & 0.959031  &  0.0641975 \\
  6&2
  &0.457142    &1.10956    &1.82857   & 2.77702    &5.14888    &1.75238    &-3.49373    &6.72161    &6.49161    &-0.0973544\\ 
  8&3  
&0.270899    &0.519396    &1.62540    &1.63882    &4.72072    &3.18307    &-2.83951    &7.57355    &13.1654    &1.94521\\
  10&4  
&0.1847    &0.2031    &1.4776    &0.7956    &3.6440    &4.3713    &-3.349    &4.9487    &17.4612    &5.5856\\
  12&5  
&0.1364    &0.0106    &1.3640    &0.2781    &2.3385    &5.3990    &    &0.9026    &18.7458    &10.4981\\
  14&6  
&0.1061    &-0.1158    &1.2730    &0.0070    &0.9553    &6.3121    &    &-3.1990    &16.9759    &16.4817\\
  16&7  
&0.0856    &-0.2033    &1.1982    &-0.0860    &-0.4423    &7.1390    &    &-6.5399    &12.2613    &23.4000\\
  18&8  
&0.0709    &-0.2660    &1.1351    &-0.0496    &-1.8261    &7.8984    &    &-8.6310    &4.7480    &31.1546
\\
 \hline
\end{tabular}
\caption{
  Perturbative coefficients for the vector correlator. The results for $n_f$ = 4 are summarized from \cite{Maier:2007yn,Maier:2009fz,Kiyo:2009gb}.
}
\label{tab:ckvec}
\end{table}

\subsection{Formulae for the extraction of $m_c$ and $\alpha_s$}
For the extraction of charm quark mass and strong coupling constant,
we impose the equality between the lattice and perturbative moments,
following the method introduced in 
\cite{Allison:2008xk,McNeile:2010ji}.
In the following we consider the pseudo-scalar channel unless
otherwise stated and suppress the superscript $PS$.

In order to reduce the discretization effects, we define the reduced
moment $R_n$ using the moment $G_n^{(0)}$ evaluated at tree level
using the same lattice formulation.
Namely,
\begin{equation}
  \label{eq:reduced_moment}
  R_n = 
  \left\{
    \begin{array}{ll}
      \displaystyle
      \frac{G_4}{G_4^{(0)}} & \mbox{for}\;\; n=4,
      \\
      \displaystyle
      \frac{am_{\eta_c}}{2a\tilde{m}_c}
      \left(\frac{G_n}{G_n^{(0)}}\right)^{1/(n-4)} 
      & \mbox{for}\;\; n\geq 6.
    \end{array}
  \right.
\end{equation}

\begin{equation}
  \label{eq:reduced_moment_V}
  R_n^V = 
    \begin{array}{ll}
      \displaystyle
      \frac{am_{J/\psi}}{2a\tilde{m}_c}
      \left(\frac{G_n^V}{G_n^{V(0)}}\right)^{1/(n-2)} 
      & \mbox{for}\;\; n\geq 4.
    \end{array}
\end{equation}
Here $m_{\eta_c}$ ($m_{J/\psi}$) represents the mass of the $\eta_c$
(${J/\psi}$) meson
calculated on the lattice, and 
$\tilde{m}_c$ is the charm quark pole mass at the tree-level 
on the same lattice ensemble. 
For domain-wall fermions, the pole mass at tree-level is given by
\be{
  \label{eq:polecharm_expand}
  a\tilde{m}_c =
  am_c\left[
    1 -\frac{1}{6}(am_c)^2 -\frac{7}{40}(am_c)^4
    -\frac{5}{112}(am_c)^6 +\frac{53}{1152}(am_c)^8+...
  \right]
}
as a function of the input quark mass $am_c$ on the lattice.
Details are in Appendix~\ref{app:polemass}.
The correction term starts at $(am_c)^2$, and its size is 
3.9\% at $am_c$ = 0.4404,
which corresponds to the input charm quark mass on our coarsest lattice.
This correction is expected to partly cancel the discretization effect
in the calculation of $am_{\eta_c}$.
Overall, in the ratios of (\ref{eq:reduced_moment}), 
the discretization effects cancel between numerator and denominator at
the leading order, {\it i.e.} $O(\alpha_s^0)$, 
and the remaining error starts at $O(\alpha_s a^2)$ for
$O(a)$-improved lattice actions.

Another definition of the reduced moment $\tilde{R}_n$ is used in \cite{Chakraborty:2014aca}:
\be{
  \label{eq:reduced_moment_latsp}
  \tilde{R}_n = 
    \begin{array}{ll}
      \displaystyle
      \frac{a}{a\tilde{m}_c}
      \left(\frac{G_n}{G_n^{(0)}}\right)^{1/(n-4)} 
      & \mbox{for}\;\; n\geq 6.
    \end{array}
}
It does not involve the meson mass $am_{\eta_{c}}$, and thus is free
from the fitting error of the correlator using the exponential function
$\m{exp}(-(am_{\eta_{c}})(t/a))$. 
On the other hand, it contains an explicit factor of the lattice spacing
$a$, and the error of the input for the lattice scale directly
reflects in the result of $m_c$.
The advantage of having the factor $m_{\eta_c}/\tilde{m}_c$ 
(or $m_{J/\psi}/\tilde{m}_c$) in (\ref{eq:reduced_moment}) 
(or in (\ref{eq:reduced_moment_V})) is that 
the meson mass $m_{\eta_c}$ (or $m_{J/\psi}$) effectively plays the
role of the input scale to determine $m_c$. 
With $\tilde{R}_n$, the error in setting the lattice spacing, which is
about 1.7\% in our case, directly appears in the final result for $m_c$.
We analyzed the data for both $R_n$ and $\tilde{R}_n$, and it turned
out that $R_n$ gives more precise determination.
Only the results with $R_n$ are presented in this paper.

On the continuum side, one defines the reduced moment $r_n$
from the derivatives of $q^2\Pi(q^2)$ with respect to $q^2$
\begin{equation}
  g_{2k}  
  \equiv
  \frac{12\pi^2Q_f^2}{k!}
  \left(\frac{\partial}{\partial z}\right)^k 
      \left.\left(z\Pi(q^2)\right)\right|_{q^2=0}
  =
  \frac{12\pi^2Q_f^2}{(k-1)!}
  \left(\frac{\partial}{\partial z}\right)^{k-1} 
  \left.\left(\Pi(q^2)\right)\right|_{q^2=0},
\end{equation}
as
\begin{equation}
  r_n
  =
  \left\{
    \begin{array}{ll}
      \displaystyle
      g_4/g_4^{(0)} = C_1/C_1^{(0)}
      & \mbox{for}\;\; n=4,
      \\
      (g_n/g_n^{(0)})^{1/(n-4)} =
      (C_{n/2-1}/C_{n/2-1}^{(0)})^{1/(n-4)}
      & \mbox{for}\;\; n\geq 6.
    \end{array}
  \right.
\end{equation}

\begin{equation}
  r_n^V
  =
    \begin{array}{ll}
      (g_n^V/g_n^{V(0)})^{1/(n-2)} =
      (C_{n/2-1}/C_{n/2-1}^{(0)})^{1/(n-2)}
      & \mbox{for}\;\; n\geq 4.
    \end{array}
\end{equation}

The tree-level moment $g_n^{(0)}$ can be explicitly written as \cite{Novikov:1977dq}
\be{
g_{2n+2}^{(0)} = 
	12\pi^2Q_f^2\frac{3}{8\pi^2} \frac{2^n(n-1)!}{(2n+1)!!},
}
and $g_{2k+2}^{V(0)}$ as
\be{
g_{2n+2}^{V(0)} = 12\pi^2Q_f^2\frac{1}{4\pi^2}\frac{2^n(n+1)(n-1)!}{(2n+3)!!}.
} 
Then, the equality (\ref{eq:G=g}) may be rewritten as
\begin{equation}
  R_n = \frac{m^{\mathrm{exp}}_{\eta_c}}{2m_c(\mu)}
  r_n(\alpha_s(\mu),m_c(\mu)).
  \label{eq:Rn=rn}
\end{equation}
Here, $r_n$ is a function of $\alpha_s(\mu)$ and $m_c(\mu)$, 
and the equation is understood as a condition to be satisfied 
by the parameters $\alpha_s(\mu)$ and $m_c(\mu)$ when a numerical
value for $R_n$ is non-perturbatively calculated on the lattice.
We can also use a ratio of the reduced moments, 
\begin{equation}
  \frac{R_n}{R_{n+2}} = \frac{
    r_n(\alpha_s(\mu),m_c(\mu))
  }{
    r_{n+2}(\alpha_s(\mu),m_c(\mu))
  },
  \label{eq:Rn=rn_ratio}
\end{equation}
which may play a complementary role to (\ref{eq:Rn=rn}), since the
truncation error of its perturbative expansion is
different from that of individual $r_n$. 

In QCD, the perturbation theory is reliable only in the relatively
short-distance regime compared to the hadronic scale 
$1/\Lambda_{\rm QCD}$.
In order to avoid the non-perturbative regime, $n$ has to be
small to satisfy a condition $n/M\ll \pi/\Lambda_{\rm QCD}$, which
implies an upper limit for $n$, 
{\it i.e.} $n\ll \pi M/\Lambda_{\rm QCD}$.
For the charmonium of $m\simeq$ 3~GeV, this means that $n$ has to be
of order of 10 or smaller.
As shown in Section~\ref{sec:Non-perturbative_corrections},
the leading non-perturbative effect 
in the Operator Product Expansion (OPE)
appears as a contribution of the gluon condensate.
Its coefficient $a_{n/2}$ in (\ref{eq:gluon_cond_pert}) rapidly grows for
larger $n$.

Combined with the lower limit for $n$ to avoid the large
discretization effect, as discussed earlier in this section,
there is a limited window of $n$ for this method to be useful.
In our analyses, we chose $n$ = 6, 8, and 10.
There is a practical limitation for $n\ge$ 12, {\it i.e} 
the $O(\alpha_s^3)$ coefficients in the perturbative expansion of
$r_n$ are not available.

\section{Lattice details}
\label{sec:lattice_details}

\begin{table}
\begin{tabular}{ccccc|cc|cc|c} 
  \hline\hline
  $\beta$ & $a^{-1}$ &
  $L^3\times T(\times L_5)$ &
   $N_{src}$ & \#meas &
  $am_{ud}$ & $am_{s}$  & 
  $m_{\pi}$ & $m_\pi L$ & id
  \\
  & [GeV] & & & & & & [MeV] & &
  \\ \hline
  4.17 & 2.453(4) &$32^3\times 64(\times 12)$ & 8 & 800& 
  0.0035 & 0.040  & 230(1) & 3.0 & C-$ud$2-$s$a\\ 
  &&&&&
  0.007	& 0.030  & 310(1) & 4.0 & C-$ud$3-$s$b\\
  &&&&&
  0.007 & 0.040  & 309(1) & 4.0 & C-$ud$3-$s$a\\
  &&&&&
  0.012	& 0.030  & 397(1) & 5.2 & C-$ud$4-$s$b\\
  &&&&&
  0.012	& 0.040	 & 399(1) & 5.2 & C-$ud$4-$s$a\\
  &&&&&
  0.019	& 0.030	 & 498(1) & 6.5 & C-$ud$5-$s$b\\
  &&&&&
  0.019	& 0.040  & 499(1) & 6.5 & C-$ud$5-$s$a\\
  &&
  $48^3 \times 96(\times 12)$ & 8 & 800 &
  0.0035 & 0.040  & 226(1) & 4.4 & C-$ud$2-$s$a-L\\
  \hline 
  4.35 & 3.610(9) & $48^3\times 96(\times 8)$ & 12 & 600 &
  0.0042 & 0.0180  & 296(1) & 3.9 & M-$ud$3-$s$b\\ 
  &&&&&
  0.0042 & 0.0250  & 300(1) & 3.9 & M-$ud$3-$s$a\\
  &&&&&
  0.0080 & 0.0180  & 407(1) & 5.4 & M-$ud$4-$s$b\\ 
  &&&&&
  0.0080 & 0.0250  & 408(1) & 5.4 & M-$ud$4-$s$a\\ 
  &&&&&
  0.0120 & 0.0180  & 499(1) & 6.6 & M-$ud$5-$s$b\\ 
  &&& &&
  0.0120 & 0.0250  & 501(1) & 6.6 & M-$ud$5-$s$a\\
  \hline
  4.47 & 4.496(9) & $64^3 \times 128(\times 8)$ & 8 & 400 &
  0.0030 & 0.015  & 284(1) & 4.0 & F-$ud$3-$s$a\\
  \hline\hline
\end{tabular}
\caption{
  Lattice ensembles used in this study.
}
\label{tab:setup}
\end{table}

We have performed a set of lattice QCD simulations with 2+1 flavors of
dynamical quarks.
The gauge action is that of tree-level Symanzik improved, and the
fermion formulation is the M\"obius domain-wall fermions
\cite{Brower:2012vk}.
The gauge links are smeared by applying the stout smearing 
\cite{Morningstar:2003gk} three times.
With this choice, the residual mass, which quantifies the violation of
the Ginsparg-Wilson relation, is under good control,
{\it i.e.} the residual mass is of $O(\mbox{1~MeV})$ on our coarsest
lattice and much smaller on finer lattices.
The effect of such a small violation can be neglected for the
charmonium correlators.
Light sea quark masses are extrapolated to the physical value such that the 
physical pion and kaon masses are reproduced.
Since the sea quark mass dependence of $R_n$ is minor, this is not a
major source of uncertainty.

There are 15 ensembles of different lattice spacings and quark masses
as listed in Table~\ref{tab:setup}.
Lattice spacings are $a$ = 0.080, 0.055, and 0.044~fm.
The spatial size of these lattices is $L/a$ = 32, 48, and 64,
respectively, to keep the physical lattice size $L$ approximately
constant, $\sim$ 2.6--2.8~fm.
The temporal size $T/a$ is always twice longer than $L/a$.
Each ensemble consists of 10,000 molecular dynamics trajectories, out
of which we chose 50-100 gauge configurations equally separated and
calculated the charmonium correlators eight or twelve times starting
from different time slices on each configuration with a $Z_2$ noise.
The number of measurement ``\#meas'' is thus 400-800 depending on
the ensemble as listed in the table.

The $Z_2$ noise is introduced to improve the statistical signal.
Namely, the $Z_2$ ($\pm 1$) noise is scattered over a time-slice as a
source to calculate the charm quark propagator; only the local $Z_2$
invariant contribution survives after averaging over the noise, so
that the desired contraction of charm and anti-charm propagators
survive and other gauge non-invariant contributions vanish.
In spite of the noise introduced, the signal is improved by
averaging over the source points.

Each ensemble has an ``id'' name, which distinguishes
coarse (C), medium (M), and fine (F) lattices,
as well as the mass of $ud$ and $s$ quark masses.
In the main ensembles (C and M), two values of strange quark mass are
chosen to sandwich the physical value from above (a) or from below (b).
On the coarse lattice at the lightest $ud$ quark mass, there is an
ensemble of larger volume of size $48^3\times 96$, which is indicated
by ``-L''.
The difference between C-$ud$2-$s$a and C-$ud$2-$s$a-L is used to
estimate the possible finite volume effect, as they have the smallest
$ud$ quark mass and the effect of finite spatial volume is expected to
be most significant in our ensembles.

The lattice spacing is set through the Wilson-flow scale $t_0$
\cite{Luscher:2010iy}.
For its physical value, we input 
$t_0^{1/2}$ = 0.1465(21)(13) fm
\cite{Borsanyi:2012zs}.
The resulting values of $a^{-1}$ are listed in Table~\ref{tab:setup}.
The table lists the central values
and the statistical error in our measurement of $t_0$.
The error in this input value is to be added for each value of $a^{-1}$.

Some details of the ensemble generation are available in
\cite{Kaneko:2013jla,Noaki:2014ura}.
The same gauge ensembles have so far been used for a calculation of
the $\eta'$ meson mass \cite{Fukaya:2015ara}, an analysis of
short-distance current correlator \cite{Tomii:2015exs}, and a
calculation of heavy-light meson decay constants 
\cite{Fahy:2015xka}.
The numerical calculations are performed using the IroIro++ code set
for lattice QCD \cite{Cossu:2013ola}.

For the vector current, we multiply the renormalization constant $Z_V$
obtained from the analysis of short-distance current correlator of
light quarks \cite{Tomii:2016xiv}.
The numerical values are 
0.9553(92) at $\beta$ = 4.17, 
0.9636(58) at $\beta$ = 4.35, and 
0.9699(47) at $\beta$ = 4.47,
where errors include statistical and systematic ones added in
quadrature. 

On each ensemble, we calculate the charmonium correlator at a bare
charm quark mass 0.4404, 0.2723 or 0.2105 at 
$\beta$ = 4.17, 4.35 and 4.47, respectively.
They are slightly mistuned to the physical charm quark mass, 
which we set by the spin-averaged mass of the 1S charmonium states
$(m_{\eta_c}+3m_{J/\psi})/4$.
We correct this minor shift by using supplemental data set taken at
three values of bare charm quark mass sandwiching the physical value.
The supplemental data are obtained with a local source and therefore 
less precise, but only used for a small interpolation of the main data
to the physical charm quark mass.

In the calculation of the charmonium correlator, we do not take
account of the contribution of disconnected quark-loop diagrams, which
may exist in the nature for the flavor-singlet operators like
$j_5=\bar{\psi}_c\gamma_5\psi_c$.
For the correspondence between the lattice and perturbative
calculations, this does not cause any problem because one can omit the
corresponding diagrams also in perturbation theory.
For the input to tune the charm quark mass on the lattice, this could
lead to some bias, as the physical input parameter, a mass of
$\eta_c$ or $J/\psi$, includes such effect.
Furthermore, the electromagnetic correction which is neglected in our
lattice calculation could also be a source of systematic error.
These sources of uncertainties are discussed in some detail in
Section~\ref{sec:determination}.

\section{Temporal moments of vector current correlator}
\label{sec:vector_moments}
As described in Section~\ref{sec:time_moments}, the temporal moments of
the charmonium vector-current correlator can be compared with the
experimental value.

Analogous to the reduced moments defined for the pseudo-scalar channel
(\ref{eq:reduced_moment}), we define the reduced moments $R_n^V$ 
for the vector moments (\ref{eq:momentV}).
We can then write the correspondence between the lattice and continuum
as 
\begin{equation}
  R_{2k+2}^V = m_{J/\psi}
  \left(\frac{M_k}{g_{2k+2}^{V(0)}}\right)^{\frac{1}{2k}},
\end{equation}
which is obtained from (\ref{eq:G_n^V}).

Numerical results for $Z_V^{-\frac{2}{n-2}}R_n^V$ are summarized in 
Table~\ref{tab:Rn_vec} for $n(=2k+2)$ = 6, 8, 10 and 12.
For each ensemble, the results are interpolated to the physical charm
quark mass; the statistical error is propagated by the bootstrap method. 

\begin{figure}[tbp]
  \includegraphics[width=8cm, angle=-90]{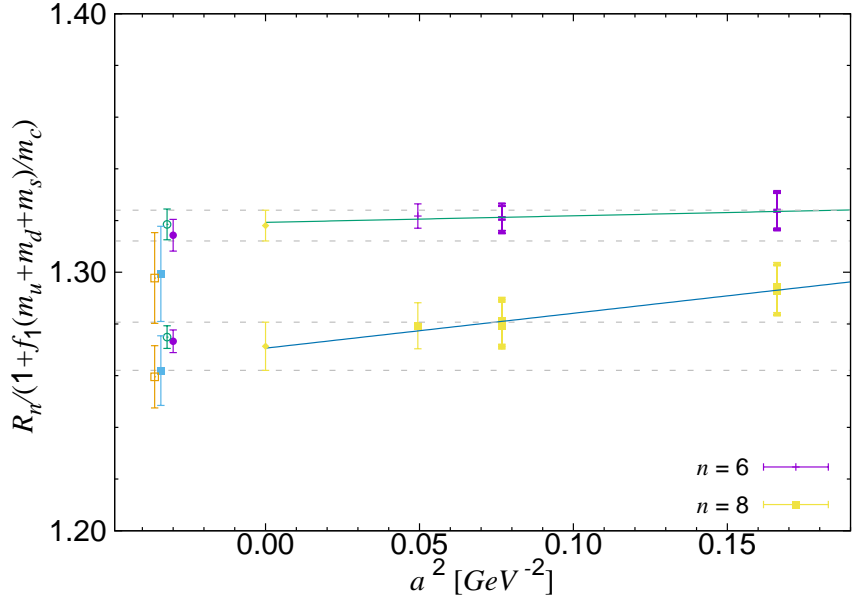}
  \caption{
    Continuum extrapolation of the reduced moments for the vector
    current $R_n^V$ ($n$ = 6 (pluses) and 8 (squares)).
    Data are plotted after correcting for the finite light quark mass
    effects by multiplying $1/(1+f_1(m_u+m_d+m_s)/m_c)$.
    Lattice data are corrected for the missing charm quark
    loop effect, estimated by perturbation theory, 
    $r_n^V(n_f=4)/r_n^V(n_f=3)$.
    The error of the individual lattice data includes that from
    the renormalization factor $Z_V$, which is the dominant source of
    error. 
    The points at $a^2=0$ are our estimate of the continuum limit
    based on two methods of continuum extrapolation.
    Its error includes that due to the input for $a^{-1}$ as well.
    Phenomenological estimates of corresponding quantities 
    are plotted on the left: 
    Dehnadi {\it et al.} \cite{Dehnadi:2011gc} (filled circle), 
    Kuhn {\it et al.} \cite{Kuhn:2007vp} (open circle), 
    Kuhn {\it et al.} \cite{Kuhn:2001dm} (filled square), 
    and Hoang {\it et al.} \cite{Hoang:2004xm} (open square).
  }
\label{fig:vectorex}
\end{figure}

The results are linearly extrapolated to the physical light quark
mass and plotted as a function of $a^2$ in Figure~\ref{fig:vectorex}.
The lattice results are nearly constant in $a^2$, and the continuum
extrapolation as discussed below is also shown.

\begin{table}
  \begin{tabular}{c|cccc} \hline
    & $Z_V^{-\frac{2}{n-2}}R_6^V$ & $Z_V^{-\frac{2}{n-2}}R_8^V$&$Z_V^{-\frac{2}{n-2}}R_{10}^V$ &$Z_V^{-\frac{2}{n-2}}R_{12}^V$ \\ \hline
    
    C-$ud$2-$s$a&	1.3563(5)&	1.3101(5)&	1.2722(5)&	1.2429(5)	\\
    C-$ud$3-$s$b&	1.3562(5)&	1.3101(5)&	1.2721(5)&	1.2428(5)	\\
    C-$ud$3-$s$a&	1.3563(5)&	1.3102(5)&	1.2722(5)&	1.2430(5)	\\
    C-$ud$4-$s$b&	1.3564(5)&	1.3103(5)&	1.2723(5)&	1.2430(5)	\\
    C-$ud$4-$s$a&	1.3576(5)&	1.3112(5)&	1.2731(5)&	1.2437(5)	\\
    C-$ud$5-$s$b&	1.3589(5)&	1.3125(5)&	1.2742(5)&	1.2448(5)	\\
    C-$ud$5-$s$a&	1.3594(5)&	1.3130(5)&	1.2747(5)&	1.2452(5)	\\
    C-$ud$2-$s$a-L&	1.3559(5)&	1.3099(4)&	1.2721(4)&	1.2432(4)	\\
    \hline
    M-$ud$3-$s$b&	1.3461(7)&	1.2919(6)&	1.2553(6)&	1.2285(6)	\\
    M-$ud$3-$s$a&	1.3475(6)&	1.2932(6)&	1.2564(6)&	1.2296(5)	\\
    M-$ud$4-$s$b&	1.3483(7)&	1.2939(6)&	1.2571(6)&	1.2302(6)	\\
    M-$ud$4-$s$a&	1.3489(6)&	1.2944(6)&	1.2575(6)&	1.2306(6)	\\
    M-$ud$5-$s$b&	1.3499(7)&	1.2953(6)&	1.2583(6)&	1.2312(6)	\\
    M-$ud$5-$s$a&	1.3511(6)&	1.2964(6)&	1.2594(6)&	1.2323(6)	\\
    \hline
    F-$ud$3-$s$a &	1.3435(7)&	1.2892(6)&	1.2536(6)&	1.2275(6)	\\
    \hline
  \end{tabular}
  \caption{
    Reduced moment $Z_V^{-\frac{2}{n-2}}R_n^V$ for each ensemble. 
  }
  \label{tab:Rn_vec}
\end{table}

We extrapolate $R_n^V$ assuming the form
\begin{equation}
  R_n^V = R_n^V(0) 
  \left( 1 + c_1(am_c) ^2 \right) \times 
  \left( 1 + f_1\frac{m_u + m_d + m_s}{m_c} \right),
\label{fittingfunc}
\end{equation}
with free parameters $R_n^V(0)$, $c_1$, and $f_1$. 
The error of $O(a^2)$ is eliminated by an extrapolation with this
form, while the effect of $O(a^4)$ still needs to be estimated. 
We attempt two continuum extrapolations assuming a linear dependence
on $a^2$ with and without the point of the coarsest lattice. 
The three-point fit yields a $\chi^2/\mathrm{dof}$ = 0.17 (0.40) for
$n$ = 6 (8).
The value of $\chi^2/\mathrm{dof}$ is slightly underestimated since
the correlated systematic error for $Z_V$ among different $\beta$
values is not taken into account.
We take the mean value of these two extrapolations as a central value,
and estimate the remaining discretization error using the deviation
from the mean value.


The quark mass dependence of non-perturbative origin, 
which is assumed to be linear in $m_u + m_d + m_s$, turned out to be tiny 
($f_1 \sim 0$), and we do not consider its higher order effects. 

Since the lattice calculation is performed with three light flavors
($n_f=3$), we estimate
the effect of charm quark loop by perturbative theory. 
Namely, we correct the lattice result of $n_f=3$ to that of $n_f=4$,
by multiplying $r_n^V(n_f=4)/r_n^V(n_f=3)$.
The perturbative coefficients are calculated to $O(\alpha_s^2)$
\cite{Maier:2007yn} and partly to $O(\alpha_s^3)$
\cite{Maier:2009fz}.
We set the number of heavy flavors $n_h=1$ (or 0) for $n_f=4$ (or 3)
to calculate the ratio $r_n^V(n_f=4)/r_n^V(n_f=3)$.
We also take account of the small difference of
$\alpha_s^{n_f=4}(\mu)$ and $\alpha_s^{n_f=3}(\mu)$ as well as
that of $m_c(\mu)$.
The correction is numerically small, {\it i.e.} 
the factor is 
0.9992(26), 1.0026(68), 1.0156(342) 
for $n$ = 6, 8, and 10, respectively.

Table~\ref{tab:R_n_vec extrap} summarizes the results for $R_n^V(0)$.
The perturbative error is estimated by taking a range of the scale
$\mu$ = 2--4~GeV.
The large error for $n=10$ is due to the lack of $O(\alpha_s^3)$
formula. 
The results in the continuum limit are compared with the
phenomenological estimates \cite{Kuhn:2001dm,Hoang:2004xm,Kuhn:2007vp,Dehnadi:2011gc,Dehnadi:2015fra}.
The agreement of the lattice data and the phenomenological estimates
is remarkable.
In particular, our data are consistent with the updated estimates with
reduced error of \cite{Kuhn:2007vp,Dehnadi:2011gc}, and the size of
total error is comparable.

\begin{table}[tbp]
  \begin{tabular}{ccccccc}
    \hline
    \multicolumn{1}{c}{}&\multicolumn{2}{c}{This work} & 
    \multicolumn{4}{c}{Phenomenological estimates}\\
    \hline
    & $n_f=3 $ & $n_f=4$ &
    \cite{Dehnadi:2011gc} & \cite{Kuhn:2007vp} &
    \cite{Kuhn:2001dm} & \cite{Hoang:2004xm}\\
    \hline
    $R_6^V {(0)}$ & 1.3191(33)(12)(4)(34) & 1.3181(33)(13)(4)(33)(34) &
    1.3143(61) & 1.3185(59) & 1.2994(184) & 1.2978(176)\\
    $R_8^V {(0)}$ & 1.2680(22)(7)(2)(28) & 1.2714(22)(8)(2)(28)(86) &
    1.2732(44) & 1.2749(44) & 1.2620(135) & 1.2596(120)\\
    $R_{10}^V {(0)}$ & 1.2365(16)(13)(0)(22) & 1.2558(16)(13)(0)(22)(423) &
    1.2439(35) & 1.2447(34) & 1.2352(104) & 1.2330(91)\\
    \hline
  \end{tabular}
  \caption{
    Reduced moments $R_n^V (0) $ extrapolated to the
    continuum limit at physical light quark masses.
    The errors in ``this work'' are from statistical, discretization,
    finite volume, and the input value of $t_0 ^{1/2}$, respectively.
    The numbers for $n_f=3$ are the lattice data with 2+1 flavors of
    dynamical quarks, while those for $n_f=4$ are after the correction
    by $r_n(n_f=4)/r_n(n_f=3)$. 
    The last error for ``$n_f=4$'' is from this perturbative
    correction factor.
    Phenomenological estimates from 
    \cite{Dehnadi:2011gc,Kuhn:2007vp,Kuhn:2001dm,Hoang:2004xm} are
    shown with the estimated error in these references.
  }
  \label{tab:R_n_vec extrap}
\end{table}

\section{Temporal moments of pseudoscalar current correlator}
\label{sec:pseudoscalar_moments}
The reduced moments $R_n$ ($n$ = 6, 8, 10 and 12)
for the pseudo-scalar channel 
obtained at each ensemble are listed in
Table~\ref{tab:R_n} and their ratios $R_n/R_{n+2}$ are in
Table~\ref{tab:R_n ratio}.

\begin{table}
  \begin{tabular}{c|cccc} \hline
    &$R_6$ & $R_8$&$R_{10}$ &$R_{12}$ \\ \hline
    C-$ud$2-$s$a &	1.4689(6)&	1.3681(5)&	1.3087(4)&	1.2679(4)	\\ 
    C-$ud$3-$s$b &	1.4696(5)&	1.3686(5)&	1.3090(4)&	1.2682(4)	\\
    C-$ud$3-$s$a &	1.4692(5)&	1.3683(5)&	1.3089(4)&	1.2681(4)	\\
    C-$ud$4-$s$b &	1.4696(6)&	1.3687(5)&	1.3091(4)&	1.2683(4)	\\
    C-$ud$4-$s$a &	1.4706(5)&	1.3693(5)&	1.3097(4)&	1.2687(4)	\\
    C-$ud$5-$s$b &	1.4720(5)&	1.3705(5)&	1.3107(4)&	1.2696(4)	\\
    C-$ud$5-$s$a &	1.4722(6)&	1.3708(5)&	1.3109(4)&	1.2699(4)	\\
    C-$ud$2-$s$a-L &	1.4693(5)&	1.3684(4)&	1.3091(4)&	1.2685(4)	\\
    \hline
    M-$ud$3-$s$b &	1.4869(6)&	1.3598(5)&	1.2977(5)&	1.2582(4)	\\
    M-$ud$3-$s$a &	1.4882(6)&	1.3609(5)&	1.2986(5)&	1.2590(4)	\\
    M-$ud$4-$s$b &	1.4888(7)&	1.3611(6)&	1.2987(5)&	1.2590(5)	\\
    M-$ud$4-$s$a &	1.4896(6)&	1.3618(5)&	1.2994(5)&	1.2596(4)	\\
    M-$ud$5-$s$b &	1.4899(7)&	1.3621(5)&	1.2996(5)&	1.2598(5)	\\
    M-$ud$5-$s$a &	1.4912(6)&	1.3631(5)&	1.3005(4)&	1.2605(4)	\\
    \hline
    F-$ud$3-$s$a &	1.4961(6)&	1.3616(5)&	1.2987(5)&	1.2590(4)	\\
    \hline
  \end{tabular}
  \caption{
    Reduced moment $R_n$ in each ensemble. 
    The errors shown are statistical.
  }
  \label{tab:R_n}
\end{table}

\begin{table}
  \begin{tabular}{c|ccc} \hline
    & $R_6/R_8$ & $R_8/R_{10}$ & $R_{10}/R_{12}$ \\ \hline
    C-$ud$2-$s$a &	1.07365(8)&	1.04540(3)&	1.03216(2)	\\
    C-$ud$3-$s$b &	1.07381(7)&	1.04548(3)&	1.03220(2)	\\
    C-$ud$3-$s$a &	1.07368(7)&	1.04542(3)&	1.03217(2)	\\
    C-$ud$4-$s$b &	1.07377(7)&	1.04547(3)&	1.03221(2)	\\
    C-$ud$4-$s$a &	1.07396(7)&	1.04555(3)&	1.03226(2)	\\
    C-$ud$5-$s$b &	1.07406(7)&	1.04563(3)&	1.03234(2)	\\
    C-$ud$5-$s$a &	1.07400(8)&	1.04563(3)&	1.03235(2)	\\
    C-$ud$2-$s$a-L &	1.07369(5)&	1.04529(2)&	1.03204(1)	\\
    \hline
    M-$ud$3-$s$b &	1.09346(11)&	1.04783(5)&	1.03144(3)	\\
    M-$ud$3-$s$a &	1.09360(9)&	1.04792(4)&	1.03151(2)	\\
    M-$ud$4-$s$b &	1.09380(12)&	1.04801(5)&	1.03157(3)	\\
    M-$ud$4-$s$a &	1.09384(12)&	1.04802(5)&	1.03158(3)	\\
    M-$ud$5-$s$b &	1.09388(12)&	1.04808(5)&	1.03162(3)	\\
    M-$ud$5-$s$a &	1.09401(9)&	1.04814(4)&	1.03167(2)	\\
    \hline
    F-$ud$3-$s$a &	1.09882(10)&	1.04839(4)&	1.03150(3)	\\
    \hline
  \end{tabular}
  \caption{Ratios of the reduced moment $R_n/R_{n+2}$ for each
    ensemble. 
    The errors represent that of statistical.
  }
  \label{tab:R_n ratio}
\end{table}

By comparing the data at two different volumes, which are available
for the coarse lattice with the lightest sea quarks ($\beta$ = 4.17,
$am_{ud}$ = 0.0035),
we observe that the results on the larger volume $48^3\times 96$ (C-$ud$2-$s$a-L) are
lower than those on $32^3\times 64$ (C-$ud$2-$s$a) by about two
standard deviations for $R_4$ and $R_6$.
For $R_8$, $R_{10}$ and $R_{12}$, on the other hand, the data at
different volumes coincide within the statistical error.
We estimate the systematic error due to finite volume effect by
taking these differences and applying them for all the other
ensembles assuming similar values for each.
This should give a conservative estimate because the finite volume
effect is expected to be significantly less for heavier sea quarks.
We note that the value of $m_\pi L$ is small ($\sim$ 3.0) only for
this ensemble (C-$ud$2-$s$a); others satisfy $m_\pi L>3.9$.
As listed in the table of systematic errors in final results (Table~\ref{nogluin2}),
the estimated error from this source is an order of magnitude smaller
than other sources, and any combined error of the finite volume effect
with other sources is negligible.

\begin{figure}[tbp]
  \includegraphics[width=8.0cm,angle=-90]{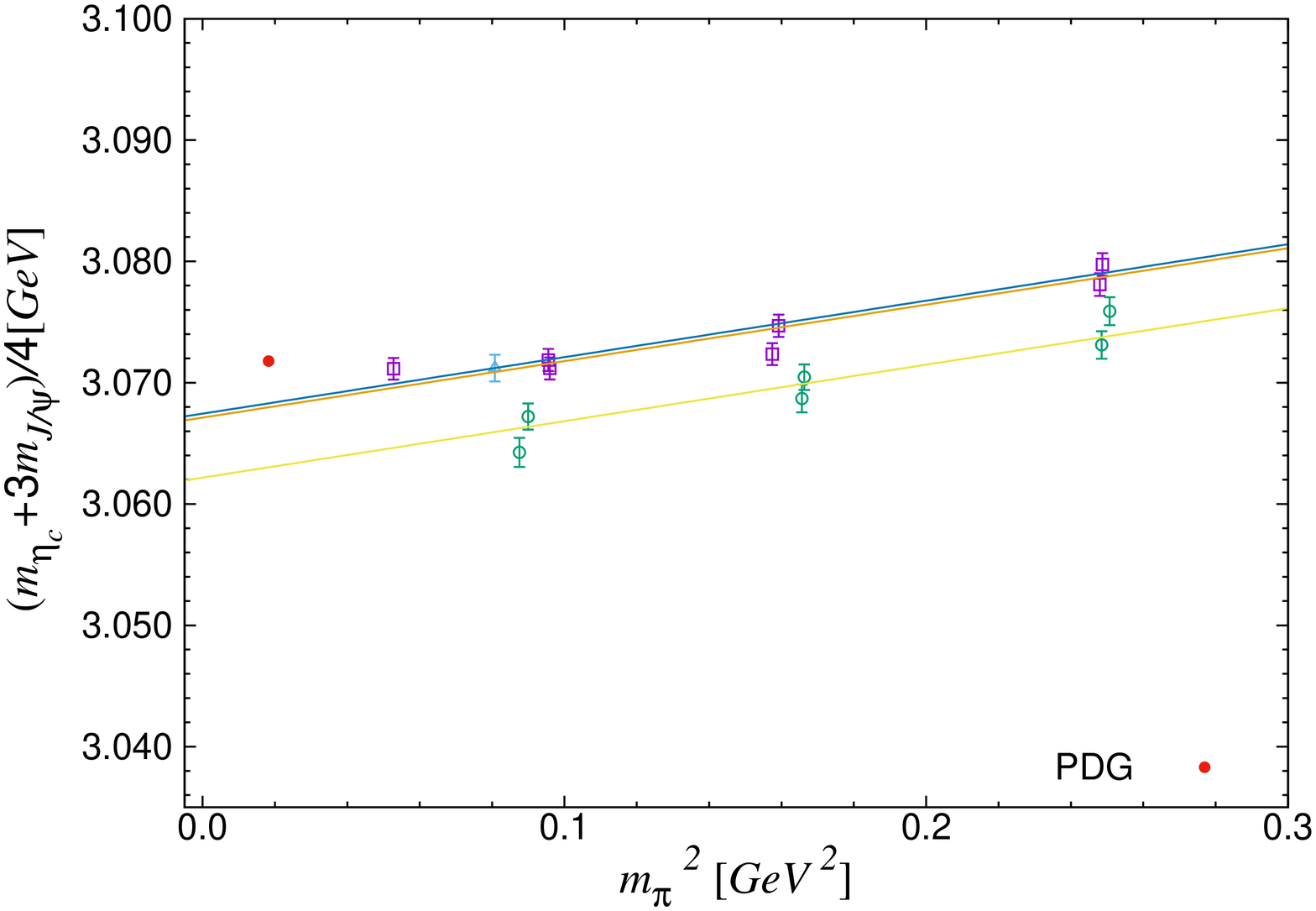}
  \caption{
    Spin-averaged mass $(m_{\eta_c}+3m_{J/\psi})/4$ as a function of $m_\pi^2$.
    The experimental value, 3.072~GeV, is shown by a filled circle.
    Data at $\beta = 4.17$ (square), $\beta = 4.35$ (circle), and
    $\beta = 4.47$ (triangle) are plotted.
    At each $\beta$, the extrapolation to the physical pion mass
    slightly misses the experimental value since the input $m_c$ is
    not exactly tuned. 
    This tiny difference is corrected when we analyze the temporal
    moments. 
  }
  \label{fig:spin_ave_phys}
\end{figure}

We interpolate $R_n$ in $m_c$ to the physical point by tuning until
the spin-averaged mass $({m_{\eta_c} + 3m_{J/\psi}})/4$ 
reproduces the experimental value, 3.0687~GeV.
Figure~\ref{fig:spin_ave_phys} shows an extrapolation of below.
the spin-averaged mass to the physical pion mass. 
A fit is done assuming that the slope in $m_\pi^2$ is independent on
$\beta$, which seems reasonable as the plot shows.
The $\chi^2/\mathrm{dof}$ of this fit is 1.9.

Our lattice results extrapolated to the physical pion mass
are slightly lower than the experimental data
by about 0.1--0.3\% depending on $\beta$ because of slight mistuning
of the input $m_c$.
We correct for them by using the supplemental data taken at three
different $m_c$'s for each $\beta$ as discussed 

\begin{figure}[tbp]
  \includegraphics[width=8.0cm,angle=-90]{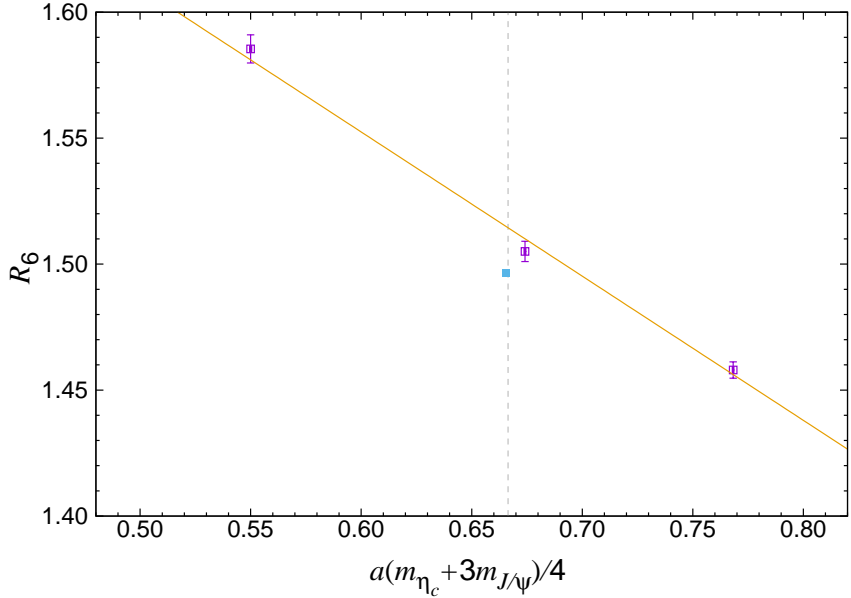}
  \caption{
    $R_6$ as a function of the spin-averaged mass 
    $a(m_{\eta_c}+3m_{J/\psi})/4$ from different $m_c$.
    Data at $\beta$ = 4.47 are shown.
    The dashed line represents the physical spin-averaged mass.
    Three data points shown by open square are from the supplemental
    data set obtained without using the $Z_2$ noise source.
    The filled square is our main data point calculated with $Z_2$
    noise source.
  }
  \label{fig:Rn_spin_ave}
\end{figure}

Figure~\ref{fig:Rn_spin_ave} is an example of the $m_c$ dependence of
$R_6$ obtained at $\beta$ = 4.47.
Our main data point (filled square) is slightly off the target
physical value of the physical $(m_{\eta_c}+3m_{J/\psi})/4$ shown by a
dashed line. 
We correct the data using a slope obtained from the supplemental data
at three values of $m_c$ shown in Figure~\ref{fig:Rn_spin_ave}.
The supplemental data have significantly larger statistical error, but
are sufficiently precise to determine the slope needed for the
correction.
(The fit to obtain the slope is uncorrelated.
The effect of ignoring the correlation among three data points should
have little impact on the final result, since the correction itself is
very small.)
The correction factor on this ensemble is tiny, {\it i.e.} 
$\sim$ 0.03\%.

When we interpolate to the physical point of $m_c$, 
we need to incorporate the uncertainty of the lattice spacing
originating from the input value of $t_0^{1/2}$.
This error is propagated to the following analysis by repeating the
same analysis with the lattice spacing $a$ set to the upper and lower
limits of its uncertainty.

\begin{figure}[tbp]
  \includegraphics[width=8.0cm,angle=-90]{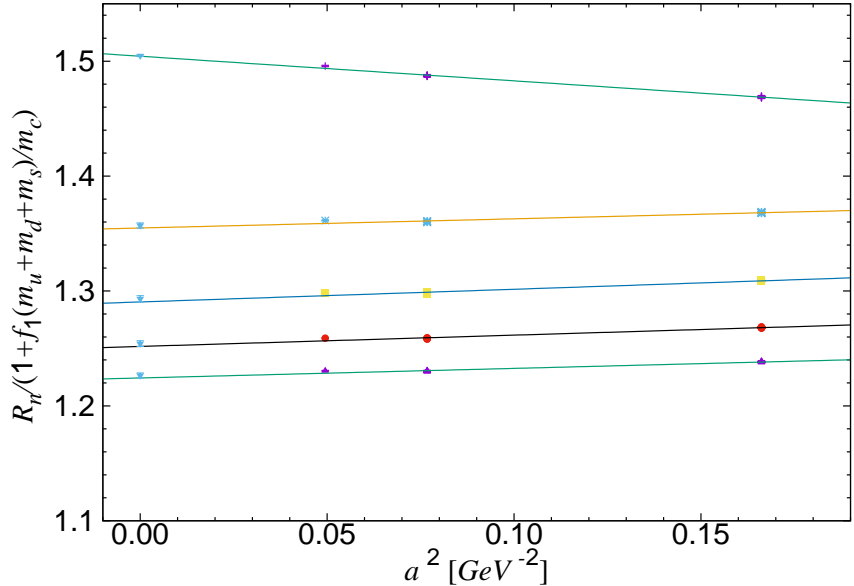}
  \caption{
    Continuum extrapolation of $R_n(a)$.
    Data points correspond to $R_6$, $R_8$, $R_{10}$, $R_{12}$, and $R_{14}$
    from top to bottom.
    The continuum extrapolation assuming the form (\ref{fitform_Rn})
    is shown by lines.
    The points at $a=0$ represent our estimate obtained from a mean of
    the extrapolated values with and without the coarsest lattice
    data. 
  }
  \label{fig:R_n extrap}
\end{figure}

We extrapolate $R_n$ to the continuum limit assuming the form similar to (\ref{fittingfunc}):
\begin{equation}
  \label{fitform_Rn}
  R_n = R_n(0) 
  \left( 1 + c_1(am_c) ^2 \right) \times 
  \left( 1 + f_1\frac{m_u + m_d + m_s}{m_c} \right).
\end{equation}
This continuum extrapolation is shown in Figure~\ref{fig:R_n extrap}.

Remaining discretization error is estimated as in the vector 
channel by taking the difference between the extrapolations with two
and three data points.
The lattice data at different values of $a$ and sea quark masses
are statistically independent.   
We use the standard $\chi ^2$ fitting; the value of
$\chi^2/\mathrm{dof}$ is
2.1, 4.1, 5.1, 4.6, and 3.9 for $R_6$, $R_8$, $R_{10}$, $R_{12}$, and
$R_{14}$, respectively.

\begin{table}[tbp]
  \begin{tabular}{ccc}
    \hline
    &$n_f$ = 3 &$n_f$ = 4 \\ \hline
    $R_6 {(0)}$&1.5048(5)(5)(4)(66)&1.5094(5)(5)(4)(66)	 \\
    $R_8 {(0)}$&1.3570(4)(22)(3)(39)&1.3589(4)(22)(3)(39)	 \\
    $R_{10}{(0)}$&1.2931(4)(27)(5)(27)&1.2965(4)(27)(5)(27)	\\
    \hline
    $R_6  {(0)}/R_8  {(0)}$&1.1089(1)(13)(0)(17) &1.1108(1)(13)(0)(17)\\
    $R_8  {(0)}/R_{10} {(0)}$&1.0494(0)(5)(1)(8)&1.0481(0)(5)(1)(8) \\
    \hline
  \end{tabular}
  \caption{
    Reduced moments $R_n$ and their ratios extrapolated to the
    continuum limit at physical light quark masses.
    The numbers for $n_f=3$ show our original calculation with $n_f =
    2+1$ on the lattice, and 
    those for $n_f=4$ are after the correction using a factor
    $r_n(n_f=4)/r_n(n_f=3)$ for $R_n$ or
    $(r_n(n_f=4)/r_n(n_f=3))/(r_{n+2}(n_f=4)/r_{n+2}(n_f=3))$ for
    $R_n/R_{n+2}$. 
    The errors represent statistical, discretization effect,
    finite volume effect, and the input value of $t_0 ^{1/2}$ in the
    order given.
  }
  \label{tab:R_n extrap}
\end{table}

Table~\ref{tab:R_n extrap} summarizes the results for $R_n(0)$. 
Systematic error due to finite volume is estimated as described above.

Again we correct the lattice result of $n_f=3$ to that of $n_f=4$, by
multiplying by $r_n(n_f=4)/r_n(n_f=3)$. 
This numerical factor is 1.0031, 1.0014, and 1.0026 for 
$n=$ 6, 8, and 10, respectively.
Table~\ref{tab:R_n extrap} lists the data before and after this
correction.

\section{Systematic errors on the continuum side}
\label{sec:matching}
As summarized in Section~\ref{sec:time_moments}, one may use
(\ref{eq:Rn=rn}) and (\ref{eq:Rn=rn_ratio}) 
to extract $\alpha_s(\mu)$
and $m_c(\mu)$ with the lattice inputs for $R_n$ obtained in the
previous section.
Several sources of systematic errors mainly on the perturbative side, are discussed in this section.

\subsection{Truncation of perturbative series}
\begin{figure}[tbp]
  \includegraphics[width=8.0cm,angle=-90]{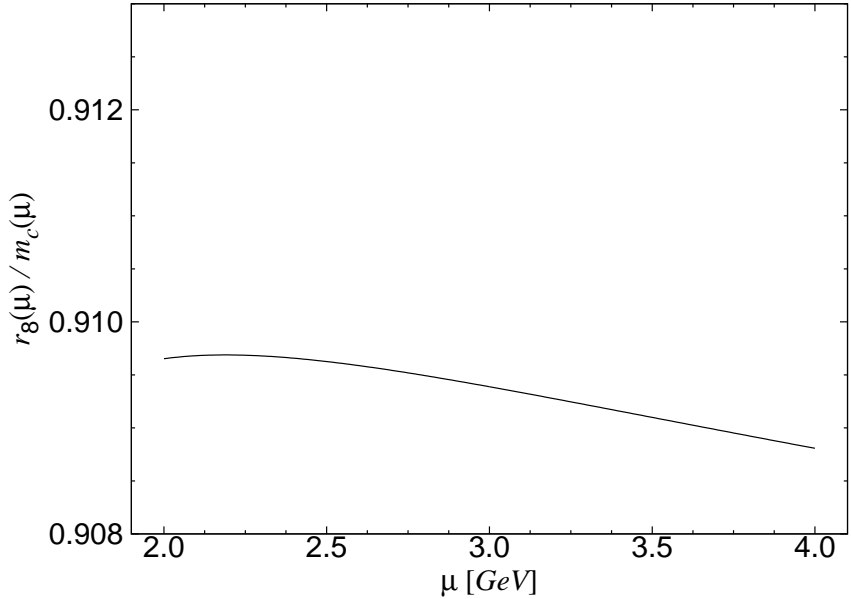}
  \caption{
    Residual scale dependence of the ratio $r_n(\mu)/m_c(\mu)$.
    The case for $n$ = 8 is plotted as a typical example.
  }
  \label{fig:mu8depend}
\end{figure}

Perturbative coefficients for $r_n$ are available up to $O(\alpha_s^3)$
as listed in Table~\ref{tab:ck}, and the remaining error is
$O(\alpha_s^4)$. 
Since the left-hand side of (\ref{eq:Rn=rn}) is independent of the
renormalization scale $\mu$, we estimate the truncation error from the
residual $\mu$ dependence of the combination 
$r_n(\alpha_s(\mu),m_c(\mu))/m_c(\mu)$
on the right-hand side.
We take $\mu$ = 3~GeV for a central value and consider the variation 
in the range of $\pm$ 1~GeV for the estimate of the truncation error.
Figure \ref{fig:mu8depend} shows an example for $n = 8$. 
The $\mu$ dependence of $r_n(\alpha_s(\mu),m_c(\mu))$ is almost
canceled by the dependence of $m_c(\mu)$, and the remnant $\mu$
dependence is tiny but non-zero which we take as the truncation error. 

We generalize this argument by taking the scale to define
$\alpha_s(\mu)$ and $m_c(\mu)$ differently.
Namely, we reorganize the perturbative series in terms of
$\alpha_s(\mu_\alpha)$ and $m_c(\mu_m)$ with $\mu_\alpha\neq\mu_m$
\cite{Dehnadi:2011gc,Dehnadi:2015fra}.
This can be done by inserting an expansion of $\alpha_s(\mu=\mu_m)$ 
in terms of $\alpha_s(\mu_\alpha)$ into the formula of
$r_n(\alpha_s(\mu),m_c(\mu))$ and rearranging the perturbative
series. 
The terms of $O(\alpha_s^4(\mu_\alpha))$ are then truncated.

After this extension, we estimate the truncation error by taking 
a variation in the range 
$\mu_\alpha=\mu_m\pm$ 1~GeV with 
$\mbox{2~GeV}\leq\mathrm{min}\{\mu_\alpha,\mu_m\}$ and $\mathrm{max}\{\mu_\alpha,\mu_m\}\leq\mathrm{4~GeV}$. 
This provides more conservative estimate of the truncation error than
simply taking $\mbox{2~GeV}\leq\mu_\alpha=\mu_m\leq\mbox{4~GeV}$.
Because of this choice, our estimate for the truncation error is
larger than those in the previous works.

\subsection{Non-perturbative corrections}
\label{sec:Non-perturbative_corrections}
The perturbative expansion is supplemented by non-perturbative
power corrections in the operator product expansion. 
Such power corrections should be carefully examined before applying
the perturbative expansion for the current correlators.

At the lowest non-trivial order, which is of the order of $1/m_c^4$,
the gluon condensate $\langle(\alpha_s/\pi)G_{\mu\nu}^2\rangle$
appears \cite{Broadhurst:1994qj}.
At the two-loop order, that is written as
\begin{equation}
  \label{eq:GG}
  \pdif{}{q^2}\left(z\Pi (q^2)^{GG}\right) = 
  \pdif{}{q^2}\left(
    \frac{\langle(\alpha _s/\pi)G_{\mu\nu}^2\rangle}{(2m_{\m{OS}})^4}
    \sum_\ell \left(a_\ell+\frac{\alpha_s}{\pi}c_\ell\right) z^\ell
  \right),
\end{equation}
where $m_{\mathrm{OS}}$ is an on-shell heavy quark mass, and
$a_\ell$ and $c_\ell$ are numerical coefficients.
The lowest order coefficients $a_\ell$ 
for the pseudoscalar (PS) and vector (V) correlators
are
\begin{equation}
  a_\ell^{PS} = -\frac{\ell-4}{12}\frac{(2)_\ell}{(3/2)_\ell},\;\;   
  a_\ell^V = -\frac{2\ell-2}{15}\frac{(4)_\ell}{(7/2)_\ell},  
\end{equation}
with $(p)_\ell$ = $\Gamma(p+\ell)/\Gamma(\ell)$.
The higher order coefficients $c_\ell$ may be found in 
\cite{Broadhurst:1994qj}.
The on-shell mass $m_{\m{OS}}$ appearing in (\ref{eq:GG}) is related
to $m_c(\mu)$ as 
$m_{\m{OS}} = m_c(\mu)[1+\alpha_s/\pi(4/3 - \log m_c(\mu)/\mu)]$
up to $O(\alpha_s^2)$ corrections.

The contribution of this term to the moment $g_{2\ell}^{GG}$ is
simply written as
\begin{equation}
  g_{2\ell}^{GG} = 
  \frac{
    \langle(\alpha_s/\pi)G_{\mu\nu}^2\rangle
  }{
    (2m_{\m{OS}})^4
  } \left(a_\ell+\frac{\alpha_s}{\pi}c_\ell\right),
\end{equation}
and the reduced moment $r_n$ is modified as
\begin{equation}
  \label{eq:gluon_cond_pert}
  r_n^{n-4} = \frac{1}{C_{n/2-1}^{(0)}}
  \left(
    C_{n/2-1} + \frac{16\pi^2}{3}
    \frac{
      \langle(\alpha_s/\pi)G_{\mu\nu}^2\rangle
    }{
      (2m_{\m{OS}})^4
    }
    \left(a_{n/2}+\frac{\alpha_s}{\pi}c_{n/2}\right)
  \right).
\end{equation}
The numerical coefficients $a_{n/2}$ are 0.179, 0.0, $-$0.208,
$-$0.449 for $n$ = 6, 8, 10, 12, respectively.

The uncertainty for the condensate 
$\langle(\alpha_s/\pi)G_{\mu\nu}^2\rangle$
is large, {\it i.e.} 
$\langle(\alpha_s/\pi)G_{\mu\nu}^2\rangle$ = 0.006$\pm$0.012 GeV$^4$ 
based on $\tau$ decay analysis \cite{Geshkenbein:2001mn}, 
or from charmonium moments 
$\langle(\alpha_s/\pi)G_{\mu\nu}^2\rangle$ = 0.005$\pm$0.004 GeV$^4$
\cite{Ioffe:2002be,Ioffe:2005ym}, 
0.022$\pm$0.004 GeV$^4$ \cite{Narison:2011xe}.
In our analysis, we treat 
$\langle(\alpha_s/\pi)G_{\mu\nu}^2\rangle$
as a free parameter and determine from the charmonium temporal
moments together with $m_c(\mu)$ and $\alpha_s(\mu)$.
Thus we avoid further uncertainty from this source.

\subsection{Effect of charm sea quark}
Our lattice simulations do not contain dynamical charm quark,
which is expected to be small since 
the leading contribution from this effect is $O(\alpha_s^2)$ and
further suppressed by a factor of $1/m_c^2$.
As already discussed, we estimate this
contribution from perturbative calculation of $r_n(n_f=4)/r_n(n_f=3)$.
We correct our lattice calculation $R_n(n_f=3)$ by multiplying this correction evaluated perturbatively at $O(\alpha_s^3)$ with $m_c(\mu=3 \m{\ GeV}) = 0.9791$ GeV, and $\alpha_s(\mu=3 \m{\ GeV}) = 0.2567$, which are taken from PDG.
The numerical factor is 1.0031, 1.0014, and 1.0026 for $n$ = 6, 8, and 10, respectively for pseudo-scalar.


\section{Determination of $m_c(\mu)$ and $\alpha_s(\mu)$}
\label{sec:determination}
We combine the non-perturbative calculation of $R_n$ with the
perturbative expansion discussed in the previous sections.

An important issue in the precise determination is that the lattice
calculation does not exactly correspond to the experimentally
observable $\eta_c$ and $J/\psi$ mesons. This is because the electromagnetic
interaction and the disconnected diagram contributions are missing.
Their masses are used to tune the charm quark mass in the lattice
calculation, and the mismatch is a potential source of systematic
error. 

Instead of including the effects of disconnect diagrams and
electromagnetic force in the lattice calculation, 
we correct the meson masses for these effects.
Namely, for the value of $m_{\eta_c}^{\m{exp}}$ in (\ref{eq:Rn=rn})
we input the experimental value 2,983.6(7) MeV after subtracting the
corrections due to disconnected and electromagnetic effects.
The effect of disconnected diagrams reduces the $\eta_c$ mass by 
2.4(8)~MeV according to a lattice study \cite{Follana:2006rc}.
The electromagnetic force is also expected to reduce the $\eta_c$ mass by
2.6(1.3)~MeV \cite{Davies:2009tsa}.

Including these potential systematic effects for the $\eta _c$ meson
mass, we use an input value
$m_{\eta_c}^{\m{exp}} = 
2983.6(0.7) + 2.4(0.8)_{\m{Disc.}} + 2.6(1.3)_{\m{EM}}$~MeV.

\begin{figure}[tbp]
  \includegraphics[width=8.0cm,angle=-90]{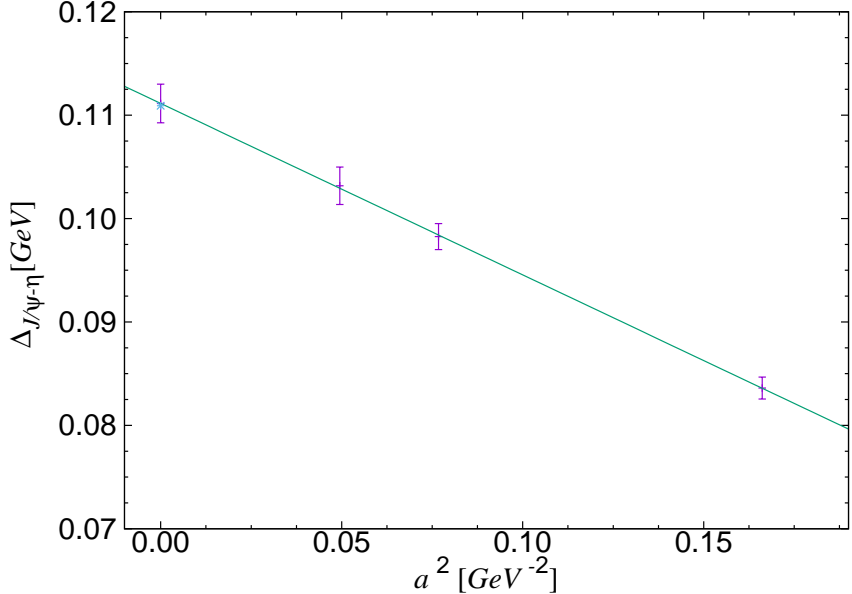}
  \caption{
    Hyperfine splitting $\Delta_{J/\psi-\eta_c}$ calculated on the
    lattice and its continuum extrapolation.
    The error of lattice scale $a$ from $t^{1/2}$ is added for
    each data point.
  }
  \label{fig:jpsietac}
\end{figure}

Discretization effect may also affect the charmonium mass spectrum
calculated on the lattice.
The hyperfine splitting 
$\Delta_{J/\psi-\eta_c}=m_{J/\psi}-m_{\eta_c}$ 
is known to be sensitive to this source of error.
Figure~\ref{fig:jpsietac} shows $\Delta_{J/\psi-\eta_c}$ as a function
of $a^2$.
A significant $a^2$ dependence is visible on the lattice data
especially for the coarsest lattice, for which the value of 
$\Delta_{J/\psi-\eta_c}$ is about 12\% lower than those at two finer
lattices.
We attempt a continuum extrapolation assuming a linear dependence on
$a^2$.
The extrapolation yields 111.4(1.8)~MeV, which is consistent with the
experimental value, 110.9(2.1)~MeV. 
It provides another evidence that the discretization effect for
the charmonium correlator is under good control 
after the extrapolation by a linear extrapolation in $a^2$.

Finally, we extract the charm quark mass $m_c(\mu)$, strong coupling
constant $\alpha_s(\mu)$, as well as the gluon condensate 
$\langle(\alpha/\pi)G^2\rangle/m^4_{\m{OS}}$, 
using (\ref{eq:Rn=rn}) 
with three temporal moments $R_6$, $R_8$ and $R_{10}$ as inputs. 
We also use the ratio of the moments $R_6/R_8$
as in (\ref{eq:Rn=rn_ratio}), which
is not independent from the individual moments but 
provides a consistency check as the truncation of perturbative
expansion is different.

\begin{figure}[tbp]
  \centering
  \includegraphics[width=10.0cm, angle=0]{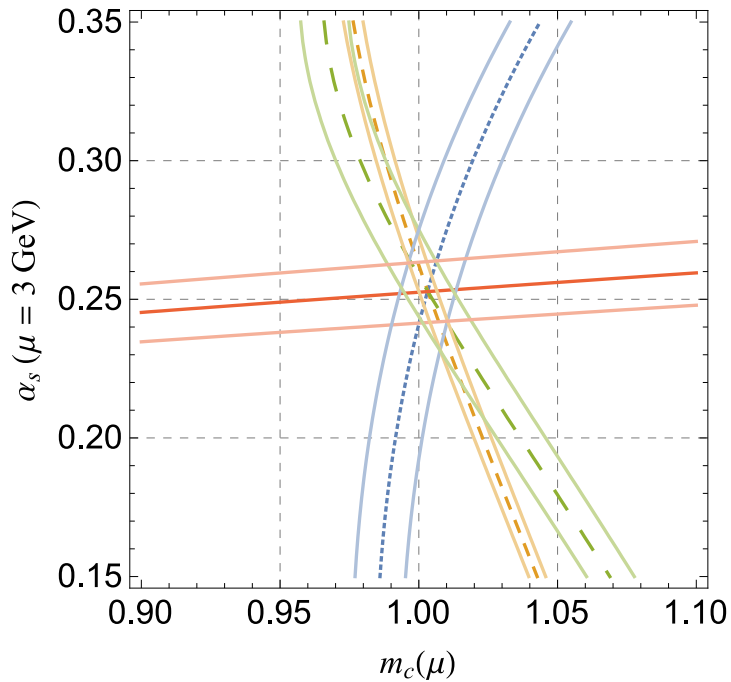}
  \caption{
    Constraints on $m_c(\mu)$ and $\alpha_s(\mu)$
    from the moments 
    $R_6$ (dotted curve), $R_8$ (dashed curve), 
    $R_{10}$ (long dashed curve), and $R_6/R_8$ (solid curve).
    For each curve, the band represents the error due to the
    truncation of perturbative expansion.
  }
  \label{fig:noingraph}
\end{figure}

Figure \ref{fig:noingraph} shows the constraints on $m_c(\mu)$ and
$\alpha_s(\mu)$ at $\mu$ = 3~GeV given by $R_6$, $R_8$, $R_{10}$, and
$R_6/R_8$.
The value of the gluon condensate is tuned such that the combination 
$R_6/R_8$, $R_8$, and $R_{10}$ give a simultaneous solution.
The plot demonstrates that each moment $R_n$ has a sensitivity to a
certain combination of $m_c(\mu)$ and $\alpha_s(\mu)$.
The ratio $R_6/R_8$, on the other hand, is sensitive only to
$\alpha_s(\mu)$, because by definition (\ref{eq:Rn=rn_ratio}) the
ratio depends on $m_c(\mu)$ only logarithmically.

\begin{table}
  \begin{tabular}{cc|cccccccc}
    \hline
    inputs & $m_c(\mu)$ [GeV] & pert &$t_0 ^{1/2}$& stat & $O(a^4)$ & vol & $m_{\eta_c}^{\m{exp}}$ & disc & EM\\
    \hline
    $R_6$, $R_8$, $R_{10}$ & 
    1.0032(98) & (82) &(51) & (5) & (16) & (4) & (3) & (4) & (6)\\
    $R_6$, $R_6/R_8$, $R_{10}$ & 
    1.0031(194) & (176) &(78) & (6) & (18) & (5) & (4) & (4) & (7)\\
    $R_6/R_8$, $R_8$, $R_{10}$ & 
    1.0033(96) & (77) &(49) & (4) & (30) & (4) & (3) & (4) & (6)\\
    \hline 
  \end{tabular}

  \smallskip
  \begin{tabular}{cc|cccccccc}
    \hline
    inputs & $\alpha_s(\mu)$ & pert &$t_0 ^{1/2}$& stat & $O(a^4)$ & vol & $m_{\eta_c}^{\m{exp}}$ & disc & EM\\
    \hline
    $R_6$, $R_8$, $R_{10}$ & 
    0.2530(256) & (213) &(134) & (12) & (38) & (10) & (9) & (10) & (16)\\
    $R_6$, $R_6/R_8$, $R_{10}$ &
    0.2528(127)	& (120)	&(33) & (2) & (25) & (1) & (0) & (0) & (1)\\
    $R_6/R_8$, $R_8$, $R_{10}$ &
    0.2528(127) & (120) &(32) & (2) & (26)	& (1) & (0) & (0) & (1)\\
    \hline
  \end{tabular}

  \smallskip
  \begin{tabular}{cc|cccccccc}
    \hline
    inputs & $\frac{<(\alpha / \pi)G^2>}{m^4}$ & pert &$t_0 ^{1/2}$& stat & $O(a^4)$ & vol & $m_{\eta_c}^{\m{exp}}$ & disc & EM\\
    \hline
    $R_6$, $R_8$, $R_{10}$ &
    $-$0.0005(99) & (85) &(45) & (4) & (23) & (4) & (3) & (4) & (6) \\
    $R_6$, $R_6/R_8$, $R_{10}$ &
    $-$0.0006(144) & (133) &(49) & (4) & (23)	& (4) & (3) & (3) & (5)\\
    $R_6/R_8$, $R_8$, $R_{10}$ &
    $-$0.0006(78) & (68) &(29) & (3) & (22) & (3) & (2) & (3) & (5)\\
    \hline 
  \end{tabular}
  \caption{
    Numerical results for $m_c(\mu)$ (top panel), $\alpha_s(\mu)$ (mid
    panel) and $\frac{<(\alpha_s/\pi)G^2>}{m^4}$ (bottom panel).
    The scale dependent quantities, $m_c(\mu)$ and $\alpha_s(\mu)$,
    are renormalized at $\mu$ = 3~GeV.
    The results are listed for different choices of three input
    quantities out of $R_6$, $R_8$, $R_{10}$ and $R_6/R_8$.
    In addition to the central values with combined errors, the
    breakdown of the error is presented.
    They are the estimated errors from the truncation of perturbative
    expansion, the input value of $t_0 ^{1/2}$, statistical, discretization error of $O(a^4)$ (or
    $O(\alpha_sa^2)$),
    finite volume, experimental data for $m_{\eta_c}^{\m{exp}}$,
    disconnected contribution, electromagnetic effect, in the order
    given. 
    The total error is estimated by adding the individual errors in
    quadrature. 
  }
  \label{nogluin2}
\end{table}

Table~\ref{nogluin2} lists the numerical results for the three
parameters including the breakdown of estimated errors.
They include those from the truncation of perturbative expansion,
statistical, discretization error of $O(a^4)$ (or $O(\alpha_sa^2)$), 
finite volume, experimental data for $m_{\eta_c}^{\m{exp}}$,
disconnected contribution, electromagnetic effect.
The estimation of these individual errors is already described in
previous sections.

Clearly, the truncation of the perturbative expansion is the dominant
source of error for all of these three quantities.
As described in the previous section, this source of error is
estimated conservatively by varying the scale $\mu_m$ and
$\mu_\alpha$ in the range between 2~GeV and 4~GeV excluding the region
that $\mu_m/\mu_\alpha$ is far away from 1.
The next largest error comes from the discretization effect estimated
by taking two or three data points in the continuum extrapolation.
Significance of other sources is not substantial, or even
negligible when the errors are added in quadrature.

The gluon condensate cannot be determined precisely.
In fact, our results are consistent with zero within estimated errors.
This is not surprising because this quantity is obtained as a small
difference between the perturbative and non-perturbative calculations.
It would strongly depend on the order of purturbative expansion.
Still, it shows a reasonable agreement with previous phenomenological
estimates 
\cite{Geshkenbein:2001mn,Ioffe:2002be,Ioffe:2005ym,Narison:2011xe}.

\begin{figure}[tbp]
  \centering
  \includegraphics[width=10.0cm, angle=0]{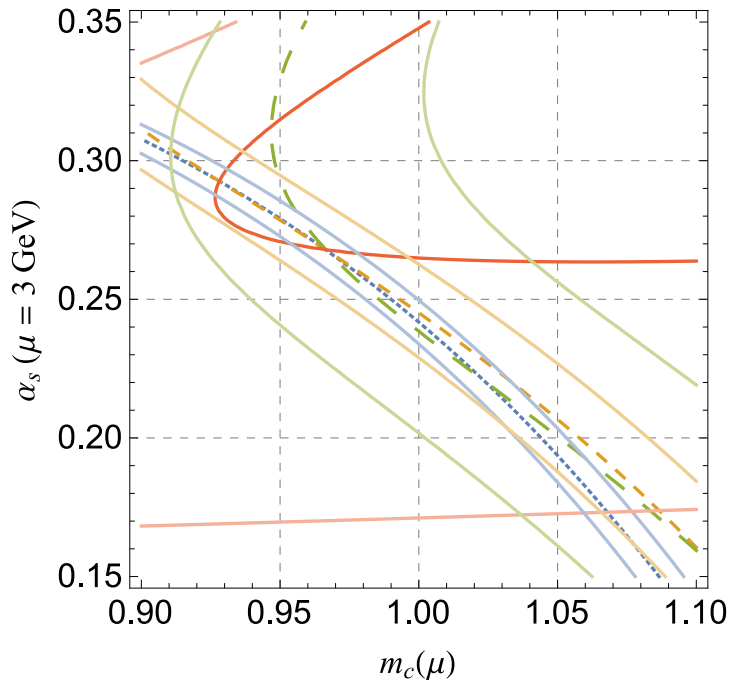}
  \caption{
    Constraint from the vector-current moments $R_n^V$ on the 
    $(m_c(\mu),\alpha_s(\mu))$ plane. 
    Dotted, dashed, long-dashed, and solid curves correspond to that
    of $R_6$, $R_8$, $R_{10}$, and $R_6/R_8$, respectively.
  }
  \label{vecplane}
\end{figure}

One may also use the vector channel to extract $m_c(\mu)$ and
$\alpha_s(\mu)$ by performing the same analysis.
Unfortunately, it was not very successful in our case.
As one can see in Figure~\ref{vecplane}, 
the constraints on the $\{m_c(\mu),\alpha_s(\mu)\}$ plane
given by different moments $R_6$, $R_8$, $R_{10}$ are
similar to each other and we are not able to disentangle $m_c(\mu)$ and $\alpha_s(\mu)$.
(The situation may be different if one can include $R_4$, but it
contains too large discretization effect as we discussed.)
Therefore, unless we use an input for $\alpha_s(\mu)$ for instance, we
are not able to use it to determine $m_c(\mu)$.
The statistical error is also 3--4 times larger for the vector
channel. 

Therefore, instead of using the vector channel to extract $m_c(\mu)$
and $\alpha_s(\mu)$, we attempt to determine $Z_V$ in (\ref{eq:G^V})
with inputs for $m_c(\mu)$ and $\alpha_s(\mu)$ obtained from the
pseudo-scalar channel.
We obtain 
0.925(19), 0.937(22) and 0.942(31) for $\beta$ = 4.17, 4.35 and 4.47, respectively.
These results are to be compared with the determination 
using the light quark hadron correlators:
0.955(9), 0.964(6), 0.970(5)
\cite{Tomii:2016xiv}.
The determination with the charm correlator is slightly lower and has
larger errors.
The ratio between the two determinations is consistent with 1, after
taking the continuum limit.

\section{Conclusion}
\label{sec:conclusion}

In this work, we extract the charm quark mass $m_c(\mu)$ and the
strong coupling constant $\alpha_s(\mu)$ through the temporal moments of
charmonium correlator calculated on lattice ensembles with 2+1 flavors
of sea light quarks described by the M\"obius domain-wall fermion.
The method was originally introduced and developed by the
HPQCD-Karlsruhe collaboration
\cite{Allison:2008xk,McNeile:2010ji,Chakraborty:2014aca},
and we apply it for the lattice data obtained with a different lattice
formulation. 

The temporal moments in the vector channel can be related to the
experimentally available moments of the spectral function, 
and provide the means to validate or to calibrate the lattice
calculations. 
For the determination of $m_c(\mu)$ and $\alpha_s(\mu)$, we 
use the pseudo-scalar channel, since the vector channel does not show
enough sensitivity to determine $m_c(\mu)$ and $\alpha_s(\mu)$
separately.

For charm quark, the discretization effect could be sizable. 
Our lattice simulations are carried out at sufficiently small lattice
spacings in the range 0.044--0.080~fm, and the continuum extrapolation
of the temporal moments is under good control.

\begin{table}
  \begin{tabular}{ccc}
    \hline
    & this work & PDG (2014)\\
    \hline
    $m_c(\mu=3\mathrm{\ GeV})$ & 1.0033(96)~GeV & \\
    $m_c(\mu=m_c)$ & 1.2871(123)~GeV & 1.275(25)~GeV\\
    $\alpha_s(\mu=3\mathrm{\ GeV})$ & 0.2528(127) & 0.2567(34)\\
    $\alpha_s(\mu=M_Z)$ & 0.1177(26) & 0.1185(6)\\
    $\Lambda_{\mathrm{\overline{MS}}}^{n_f=4}$ & 286(37)~MeV & 297(8)~MeV\\
    $\Lambda_{\mathrm{\overline{MS}}}^{n_f=5}$ & 205(32)~MeV & 214(7)~MeV\\
    \hline
  \end{tabular}
  \caption{
    Comparison of our results with the values in the 
    Review of Particle Properties (2014) \cite{Agashe:2014kda}.
    All the quantities are understood to be given in the
    $\mathrm{\overline{MS}}$ scheme. 
  }
  \label{Summary Table}
\end{table}

\begin{figure}[tbp]
  \centering
  \includegraphics[width=10.0cm, angle=-90]{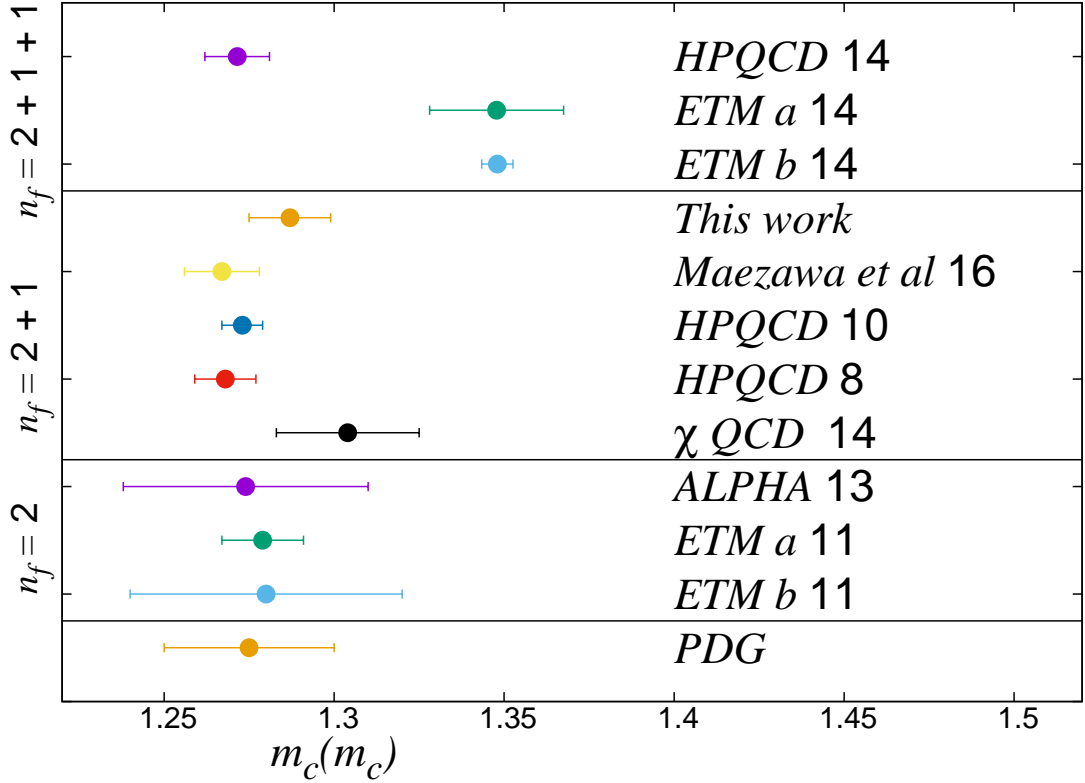}
\vspace{5truemm}
  \caption{
    Charm quark mass obtained in this work is compared with previous 
    lattice determination.
    The previous results are separately shown for different number of
    sea quarks.
    HPQCD 14 \cite{Chakraborty:2014aca}, 
    ETM $a$ 14 \cite{Alexandrou:2014sha} and 
    ETM $b$ 14 \cite{Carrasco:2014cwa} for $n_f$ = 2 + 1 + 1, 
    Maezawa {\it et al.} 16 \cite{Maezawa:2016vgv}, 
    HPQCD 10 \cite{McNeile:2010ji}, 
    HPQCD 08 \cite{Allison:2008xk}, and 
    $\chi$QCD 14 \cite{Yang:2014sea} for 
    $n_f$ = 2+1, and  
    ALPHA 13 \cite{Heitger:2013oaa}, 
    ETM $a$ 11 \cite{Jansen:2011vr}, and 
    ETM $b$ 11 \cite{Blossier:2010cr} for $n_f$ = 2.
  }
  \label{fig:compare}
\end{figure}

Our final results are compared with the PDG numbers
\cite{Agashe:2014kda} in Table~\ref{Summary Table}
and a comparison with other collaborations are shown in Figure~\ref{fig:compare}.
For our results, we take the values of the smallest total
uncertainties from Table~\ref{nogluin2}.
The charm quark mass $m_c(\mu=\mathrm{3~GeV})$ is converted to 
$m_c(\mu=m_c)$, and the strong coupling constant
$\alpha_s(\mathrm{3~GeV})$ is converted to the value at the $Z$ boson
mass using four-loop running formulae.
The threshold effect at the bottom quark mass is incorporated at
one-loop.
The resulting value of $\alpha_s(M_Z)$ is consistent with the PDG.

The result of the HPQCD collaboration \cite{Chakraborty:2014aca} for
the charm quark mass is
$m_c(\mathrm{3~GeV})$ = 0.9851(63)~GeV.
Our result is 1.8$\pm$1.2\% higher.
Since the perturbative part of the method is common, a part of the
error may be correlated among us.
 
Among various sources of the systematic error, the dominant one is the
truncation of perturbative expansion, which is currently known up to $O(\alpha_s^3)$.
In order to improve the precision on $m_c(\mu)$, therefore, higher
order perturbative calculation has a primary importance, as well as
the reduction of the scale uncertainty, which is common for all dimensionful parameters.

\begin{acknowledgments}
  We thank the members of the JLQCD collaboration.
  This work is  part of its research programs.
  Numerical simulations are performed on Hitachi SR16000 and IBM
  System Blue Gene Solution (Blue Gene/Q) at KEK under a support of
  its Large Scale Simulation Program (No. 13/14-04, 14/15-10,
  15/16-09).
  This work is supported in part by the Grant-in-Aid of the Japanese
  Ministry of Education (No. 26247043) 
  and by the Post-K supercomputer project through JICFuS.
\end{acknowledgments}

\appendix
\section{Tree-level pole mass of domain-wall fermion}
\label{app:polemass}
At the tree-level, the propagator of domain-wall fermion formulation
on the lattice is written as \cite{Aoki:2002iq}
\be{
  \label{eq:propagator}
  \langle q(-p)q(p)\rangle=
  \frac{-i\gamma_\mu \sin{p_\mu} + m(1 - W\exp{-\alpha})}
  {-(1-W\exp{\alpha}) + m^2(1 - W\exp{-\alpha})},
}
where the Wilson term $W(p)$ is
\be{
  W(p) = 1 - M - r \summation{\mu}{}(1 - \cos{p_\mu}).
}
We take the parameters $M = 1$ and $r = -1$,
according to the choice adopted in our simulations.

We obtain the pole mass at the tree-level $\tilde{m}_1$ 
by finding a pole of $\langle q(-p)q(p)\rangle$.
For zero spatial momentum, we solve the equation to define the pole
with $p_0=i\tilde{m}_1$.
The result is 
\be{
  \tilde{m}_1 = \cosh^{-1}\left(\frac{1-Q + \sqrt{3Q+Q^2}}{2}\right)
}
with $Q=((1+m^2)/(1-m^2))^2$.

\end{document}